Title: **Impact bombardment chronology of the terrestrial planets from 4.5 Ga to 3.5 Ga.**

Authors: R. Brasser[1], S. C. Werner[2] and S. J. Mojzsis[3,4]

Affiliations:

[1]Earth Life Science Institute, Tokyo Institute of Technology, Ookayama, Meguro-ku, Tokyo 152-8550, Japan.

[2]Centre for Earth Evolution and Dynamics, University of Oslo, N-0315 Oslo, Norway

[3]Department of Geological Sciences, University of Colorado, Boulder, CO 80309, USA

[4]Institute for Geological and Geochemical Research, Research Centre for Astronomy and Earth Sciences, Hungarian Academy of Sciences, H-1112 Budapest, Hungary.

**Abstract**

Subsequent to the Moon's formation, late accretion to the terrestrial planets strongly modified the physical and chemical nature of their silicate crusts and mantles. Here, we combine dynamical N-body and Monte Carlo simulations to determine impact probabilities, impact velocities, and expected mass augmentation onto the terrestrial planets from three sources: planetesimals left over from primary accretion, asteroids from the hypothetical E-belt, and comets arriving from the outer Solar System. We present new estimates of the amount of cometary material striking the terrestrial planets in an early (ca. 4480 Ma) episode of planetesimal-driven giant planet migration (Mojzsis et al., 2019). We conclude that the Moon and Mars suffer proportionally higher cometary accretion than Venus and Earth. We further conclude that the background mass addition from small leftover planetesimals to Earth and Mars is far less than independent estimates based on their respective mantle abundances of highly-siderophile elements and terrestrial tungsten isotopes. This supports the theory that both planets were struck by single large bodies that delivered most of their terminal mass augmentation since primary accretion, rather than a throng of smaller impactors. Our calculated lunar, martian and mercurian chronologies use the impacts recorded onto the planets from dynamical simulations rather than relying on the decline of the population as a whole. We present fits to the impact chronologies valid from 4500 Ma to ca. 3700 Ma by which time the low number of planetesimals remaining in the dynamical simulations causes the impact rate to drop artificially. The lunar timeline obtained from these dynamical simulations using nominal values for the masses of each contributing reservoir is at odds with both the calibrated Neukum (Neukum et al., 2001) and Werner (Werner et al., 2014; Werner, 2019) chronologies. For Mars, the match with its calibrated Werner chronology is no better; by increasing the mass of the E-belt by a factor of four the dynamical lunar and martian chronologies are in line with that of Werner (2019) and match constraints from the current population of Hungaria asteroids. Yet, neither of our dynamical timelines fit well with that of Neukum. The dynamical lunar and martian chronologies are also different from each other. Consequently, the usual extrapolation of such chronologies from one planetary body to the other is technically inappropriate.

# 1 Introduction

In traditional dynamical models, the terrestrial planets grow from a coagulation of planetesimals into protoplanets, which continue to remain submerged in a swarm of planetesimals. This system eventually evolves into a giant impact phase, wherein the protoplanets collide with the planetesimals and with each other, which ultimately leads to the terrestrial planets. There exist many different dynamical simulations to explain this history. The principal models are the classical (Chambers & Wetherill, 1998; Chambers, 2001), Grand Tack (Walsh et al., 2011; O'Brien et al., 2014; Brasser et al., 2016a), depleted disk (Izidoro et al., 2014, 2015; Raymond & Izidoro, 2017), and pebble accretion (Ormel & Klahr, 2010; Lambrechts & Johansen, 2012; Johansen & Lambrechts, 2017). In the latter, sunward-drifting sub-meter-sized objects termed "pebbles" accrete onto large planetesimals or small planetary embryos which then bifurcate and rapidly grow into either the giant planets (Levison et al., 2015a) or the terrestrial planets (Levison et al., 2015b).

A variation on the classical model invokes the formation of a narrow annulus (Hansen, 2009; Walsh & Levison, 2016). Within the annulus model one option calls for drifting pebbles to pile up in the terrestrial planet region to first form planetesimals that subsequently give rise to the terrestrial planets (Drążkowska et al., 2016). All of these models can reproduce the mass-orbital distance relation of the terrestrial planets and their orbital excitation with various degrees of success. When tied to the documented compositions of the sampled Solar System (Earth, Moon, Mars and asteroids), some are also deemed better than others at reproducing the isotopic composition of the terrestrial planets (Brasser et al., 2017, 2018; Woo et al., 2018).

Another key feature that all of the various dynamical models share is that the terrestrial planets experience a protracted history of late accretion subsequent to planet formation. That is, after core formation and initial separation of the silicate reservoirs (crust/mantle), left-over planetesimals on planet-crossing orbits are consumed by the terrestrial bodies as substantial mass supplements (Wetherill, 1977; Day et al., 2012). This takes the form of impact bombardment by comets (Gomes et al., 2015), planetesimals remaining from primary accretion (Bottke et al., 2007), and asteroids from the main belt (Minton & Malhotra, 2010; Nesvorný et al., 2017a) and the E-belt (Bottke et al., 2012). This late accretion thermally, structurally and chemically modified solid surfaces, and is widely regarded as having modulated life's emergence on the Hadean Earth (Mojzsis et al., 1996, 2019; Abramov & Mojzsis, 2009).

Evidence for such late mass additions is inferred from abundances of highly siderophile elements (HSEs; Re, Os, Ir, Ru, Pt, Rh, Pd, Au) in the mantles of Earth, the Moon and Mars, and asteroid Vesta (Dale et al., 2012; Day et al., 2012). The proportions are chondritic relative to each other, but greatly exceed their expected abundance after metal-silicate partitioning (Kimura et al., 1974). This observation indicates that the HSEs were delivered after core formation and silicate differentiation as a late veneer (Chou, 1978). An alternative theory, wherein the HSEs were left behind by inefficient metal–silicate partitioning at higher pressures and temperatures (Murthy, 1991; Righter et al., 2015; Rubie et al., 2016) has also been proposed. In particular, Rubie et al. (2016) notes that the partition coefficients of some HSEs at high pressure reported in Mann et al. (2012) cannot explain the HSE abundances without late accretion in the form of a late veneer, i.e. late accretion that was added after core closure. If HSE depletion was solely due to variably inefficient core formation, then it is unexpected that the HSE abundance pattern follows chondritic proportions. We emphasize that the fact that these proportions are chondritic is entirely consistent with a late mass addition scenario and not core formation at high pressure, but neither does it prove that this is what actually happened. As such, two options are available to interpret the HSE data: either low pressure core formation prevailed, which seems in turn to be violated by moderately siderophile element data in mantle samples (Becker et al., 2006; Mann et al., 2012), or the "Hadean Matte" hypothesis of O'Neill (1991) is valid, wherein all pre-late veneer sulfide in the Earth dragged the HSEs into the core. In either case, a late mass addition to deliver the HSEs observed in Earth's mantle today is still required.

If it is assumed that all the planetesimals have the same size and are far more numerous than the planets, then the relative amount of mass delivered to the terrestrial planets by late accretion should scale as the ratio of their gravitational cross sections and the number density and orbital distribution of objects that cross the orbit of each planet. For the Earth and the Moon, the ratio of their gravitational cross sections is approximately 20:1, with little variation with lunar distance to the Earth (see Section 2.2). Instead, HSE abundance in the terrestrial mantle and that deduced from lunar samples show that Earth experienced almost three orders of magnitude more late accretion of chondritic material than the Moon. This late mass augmentation to the planets amounts to about 0.7 wt.% for the Earth versus 0.025 wt.% for the Moon (Day et al., 2007; Day & Walker, 2015; Day et al., 2016), which is at odds with dynamical theory if it is assumed that most mass was delivered in the form of small planetesimals. A separate line of investigation using tungsten isotopes concludes that the measured $\varepsilon^{182}W$ values in the terrestrial mantle when compared to lunar values is consistent with the Earth having absorbed 0.3-0.8 wt.% of chondritic material after lunar formation that was then slowly mixed into the mantle (Willbold et al., 2015). Using the same isotopic system, Touboul et al. (2015) reported that the Moon accreted only 0.05 wt.% after its formation, somewhat more than that inferred from the lunar HSE abundance (Day & Walker, 2015). An alternative interpretation of the lunar $\varepsilon^{182}W$ values instead suggests that they track the earlier formation of the Moon (Thiemens et al., 2019). The implication of that study for lunar late accretion is that the Moon could have accreted far less, or that we need to rely on its HSE record with all its complications (e.g. Morbidelli et al., 2018; Zhu et al., 2019).

The high ratio of the terrestrial and lunar HSE mantle budgets led Bottke et al. (2010) to conclude that the size-frequency distribution of the remaining planetesimals from planet formation had to have been shallow even at large diameters (D>500 km, cumulative slope ~2). Furthermore, the majority of the mass delivered to the Earth came from a few large objects that yet lurked within the inner Solar System. When imposing a cumulative slope of the size-frequency distribution of these objects close to 2, Bottke et al. (2010) further conclude that the best-case model results yield mean and median diameters of the largest Earth and Moon impactors of 2500 to 3000 km and 250 to 300 km, respectively. The low population number of these large objects leads to a stochastic impact regime that statistically favors collisions with Earth due to the low number of available objects (Sleep et al., 1989). Brasser et al. (2016b) conclude that the Earth's HSE budget was delivered via collision with a single lunar-sized object, dubbed *Moneta* by Genda et al. (2017).

Debate also prevails over the amount of late accretion delivered to Mars. Osmium isotopes and more recent martian HSE measurements place its late accretion fraction close to 0.8 wt.% (Brandon et al., 2012; Day et al., 2016; cf. Righter et al., 2015). Recently, Woo et al. (2019) modelled the thermal, mechanical and chemical consequences of a Ceres-sized impactor they named *Nerio* to account for Mars' HSE complement from a postulated late veneer.

Even if the HSE abundance in the terrestrial and martian mantles is attributable to imperfect core formation, we still expect the Earth and Mars to undergo late accretion at least comparable to that accreted by the Moon.

Another area of general disagreement concerns the role of comets in the history of late accretionary bombardment of the terrestrial planets. It is accepted that leftover planetesimals from planetary accretion and the asteroid belt are the dominant contributions to the lunar and martian crater populations (Neukum & Ivanov 1994; Werner et al., 2002; Strom et al., 2005; Morbidelli et al., 2018). The contribution from comets, on the other hand, remains tentative (Strom et al., 2005; Rickman et al., 2017), and it has been suggested that the cometary contribution to impact cratering is minor for the Moon (Werner et al., 2002).

The timing of the cometary contribution is also contested. Based on $^{40-39}Ar$ ages of lunar materials it has been suggested that the purported late lunar cataclysm, or more commonly Late Heavy Bombardment (LHB; Tera et al., 1974; Ryder, 1990) was caused by the delivery of comets to the terrestrial planets from the outer Solar System (Gomes et al., 2005). On the other hand, recent meta-data analyses of the same $^{40-39}Ar$ ages of lunar samples indicates that they are inconclusive regarding the existence of the LHB (Boehnke & Harrison, 2016). Asteroidal and meteoritic reset ages of various chronometers with different closure temperature point to the delivery of the cometary material to the terrestrial planets most likely at or before 4.48 Ga (Mojzsis et al., 2019); a recent dynamical model of binary Trojan asteroid stability reaches a similar conclusion (Nesvorný et al., 2019). The arrival of cometary material around 4.48-4.45 Ga could naturally explain the absence of craters on the terrestrial planets attributable to comets because it is likely that their surfaces had not yet completely solidified (Mojzsis et al., 2019).

In this work, we extend the analyses of Brasser et al. (2016b), Brasser & Mojzsis (2017) and Mojzsis et al. (2019) by focusing our investigations of the impact bombardment on the terrestrial planets for the first billion years after the formation of the Moon. We have performed new dynamical simulations of leftover planetesimals, comets and asteroids to determine what mass fraction of each is accreted by the different terrestrial planets and on what timescale. We have chosen to approach the problem with the simplest dynamical model we can think of. Apart from the cometary contribution we do not include giant planet migration. In addition, we have chosen to use the simplest power-law size-frequency distribution of planetesimals that impact the planets to compare with the cratering record. The outcomes of the present work includes an analysis of the implications for our contemporary understanding of late accretion and how it affects our understanding of the crater chronologies of the inner planets.

## 2 Dynamical N-body simulations: Methodology and initial conditions

This section describes how late accretion to the terrestrial planets consists primarily of three components: a contribution from leftover planetesimals from terrestrial planet formation, a component from the asteroid belt and from comets arriving from the outer Solar System beyond Jupiter. The asteroid belt component is thought to come mostly from the proposed E-belt (Bottke et al., 2012), at least during the first billion years. Nesvorný et al. (2017a) performed a study on the long-term impact flux from the E-belt and main asteroid belt. Here we first discuss the dynamical setup that determines the amount of late accretion onto the Moon and the terrestrial planets from leftover planetesimals, the E-belt and the main asteroid belt. The contribution from comets is discussed separately owing to the fact that the dynamical setup is much more involved. We refer the reader to Mojzsis et al. (2019) for the thermal consequences of this late accretion scenario to the early Hadean Earth.

### 2.1 Leftover planetesimals

In Brasser et al. (2016b) we constrained the mass in leftover planetesimals based on the abundance of highly siderophile elements inferred for the lunar mantle. The lunar HSE abundance implies that the Moon had to have accreted a further 0.025 wt.% of chondritic material after its formation while it may have still been in a purported magma ocean state (Day & Walker, 2015). The computed lifetime of the lunar magma ocean varies greatly from study to study (Nemchin et al., 2009; Boyet & Carlson, 2007; Borg et al., 2011; Gaffney & Borg, 2014, Elkins-Tanton et al., 2011; Kamata et al., 2015); it could be as long as 200 Myr (Chen & Nimmo 2016). The precise age of lunar formation is debated, but may be close to 4.51 Ga (Barboni et al., 2017; cf. Connelly & Bizzarro, 2016). Thus, we start our dynamical simulations at the time of the formation of the Moon, which we assign at 4.5 Ga.

Our starting point is an available database of terrestrial planet formation simulations (Brasser et al., 2016a) in the framework of the Grand Tack model (Walsh et al., 2011). We took a snapshot of the planetesimal population 60 Myr after the start of the Grand Tack simulations, which approximately coincides with 4.5 Ga. From this we obtained 34374 planetesimals that had a perihelion distance q<1.5 au (so that they crossed Mars) and an aphelion distance Q<4.5 au (so that they did not cross Jupiter). The total planetesimal population of 36824 with Q<4.5 au from the Grand Tack simulations contained 2450 planetesimals on non-Mars crossing orbits (q>1.5 au), but their general orbital distribution is inconsistent with that of the asteroid belt. The effect of the non-Mars crossers is forthcoming in a follow-up study which uses an improved initial planetesimal distribution.

Our planet-crossing population was divided into 32 simulations of 1074 massless test particles each. We verified that the orbital distribution of the test particles in each simulation is representative of the whole sample. All the major planets were added: Mercury to Neptune plus dwarf planets Vesta and Ceres on their current orbits. We realise that this sudden inclusion of planets is different from those that existed in the Grand Tack simulations, which gravitationally shocks the test particles. None of the Grand

Tack simulations, however, reproduced the terrestrial planets exactly as they are today and we are not aware of an alternative approach for generating the initial conditions for the simulations.

Simulations were run with SWIFT RMVS3 (Levison & Duncan, 1994) for a model time of 0.5 Gyr with a time step of 0.01 yr. Planetesimals were removed once they were further than 40 AU from the Sun (whether bound or unbound) or when they collided with a planet or with the Sun.

**2.2 The E-belt**

The E-belt is an hypothesised now-empty extension of the asteroid belt, with its postulated inner edge just beyond the orbit of Mars (Bottke et al., 2012). Only 61 out of 36824, or 0.16%, of the leftover planetesimals from our Grand Tack simulations are in the E-belt, which is far too low a number to study its effect on the bombardment of the terrestrial planets. Thus, it makes sense to follow a different procedure.

We compute the contribution to late accretion from the E-belt by running 32 dynamical simulations of the evolution of this reservoir in the presence of the gravitational influence of the Sun and the eight major planets on their current orbits, as well as the dwarf planets Vesta and Ceres. Each simulation initially has 512 massless test particles. We employed the same initial conditions from Bottke et al. (2012): the test particles have a uniform distribution in semi-major axes *a* from 1.7–2.1 au and a main-asteroid-belt-like Gaussian distribution in eccentricity *e* and inclination *i*, with mean values of 0.15 and 8.5°, respectively, and standard deviations of 0.07 and 7°, respectively. None of them are initially on Mars-crossing orbits. The other three angles are chosen in a uniformly random fashion between 0º and 360º. Each simulation has a different realisation of the random variables but they are statistically identical. These simulations were run for a model time of 1 Gyr with SWIFT RMVS 3 with the same time step and removal criteria as the leftover planetesimals scheme cited above.

For both the E-belt and the leftover planetesimal simulations the impact probabilities with the planets were independently obtained from both counting direct impacts onto the planets and from employing the Wetherill (1967) post-process averaging procedure. Neither Vesta nor Ceres were found to have suffered actual impacts due to their small diameters so that this is only option to get their impact probabilities. The impact probability with the Moon can be scaled from that of the Earth by dividing it by 20; this method serves as an independent verification of our approach. We previously tested the computational method of Wetherill (1967) against counting direct impacts with the planets and find that the two results match well (Brasser et al., 2016b). The same is true for the simulations performed here. It was assumed that the Moon had its current distance to the Earth. While not correct, placing the Moon closer to the Earth does not dramatically change the outcome for the following reason.

The ratio of the impact probability of a planetesimal with the Moon versus the Earth is given by $\frac{P_M}{P_E} = \frac{R_M^2}{a_M^2}\frac{N(>r)}{N(>a_M)}$. This equation states that the relative impact rate onto the Moon vs. the Earth is approximately the ratio of the surface area of the satellite to the surface area of the sphere encompassed by its orbit, multiplied by the factor that accounts for the cumulative perigee distribution of the planetesimals influenced by the Earth's gravity. Here $R_M$ is the radius of the Moon, $a_M$ is the lunar semi-major axis and $\frac{N(>r)}{N(>a_M)}$ is the normalised fraction of comets passing within a distance *r* to the Earth to the total crossing the Moon's orbit (Zahnle et al., 1998); it neglects gravitational focusing by the Moon. This second fraction is given by $\frac{N(>r)}{N(>a_M)} = \frac{v_\infty^2\left(\frac{r}{a_M}\right)^2 + 2v_K^2\left(\frac{r}{a_M}\right)}{v_\infty^2 + 2v_K^2}$, where $v_K$ is the circular orbital velocity of the Moon and $v_\infty$ is the velocity of the planetesimal far from the Earth-Moon system. For the Earth $v_K = 8(R_E/a_M)^{-1/2}$ km s$^{-1}$, where $R_E$ is the radius of the Earth, and typically $v_\infty \geq 16$ km s$^{-1}$ so that generally $v_\infty \gg v_K$ and the cumulative perigee distribution is almost quadratic (a best fit yields a power of 1.9).

Hence, the product $\frac{P_M}{P_E} \approx \frac{R_M^2}{R_E^2}$ for all $a_M$ and the variation in impact probability with lunar semi-major axis is insignificant (though the impact velocity does vary by up to 20%).

Further test simulations were run where massless particles are distributed in the main asteroid belt; results show that the impact probability of objects from the main asteroid belt with the Moon and the terrestrial planets is at least a factor of 30 lower than that of the E-belt, consistent with earlier estimates from Morbidelli et al. (2010). Since the two reservoirs are thought to initially have a similar mass (Bottke et al., 2012) the contribution from the asteroid belt is insignificant during the first billion years. A detailed analysis of the terrestrial bombardment from the main asteroid belt was presented by Nesvorný et al. (2017a).

### 2.3 The main asteroid belt (A-belt)

We compute the contribution to late accretion from the A-belt by running 80 dynamical simulations of the evolution of this reservoir in the presence of the gravitational influence of the Sun and the eight major planets on their current orbits, as well as the dwarf planets Vesta and Ceres. Each simulation initially has 512 massless test particles. We employed similar initial conditions as Morbidelli et al. (2010): the test particles have a uniform distribution in semi-major axes *a* from 2.1–4.5 au, and a uniform distribution in eccentricity *e* and inclination *i*, with maximum values of 0.5 and 20°, respectively. We further impose that q>1.5 au and Q<4.5 au so that none of them are initially on Mars- of Jupiter-crossing orbits. The other three angles are chosen in a uniformly random fashion between 0º and 360º. Each simulation has a different realisation of the random variables but they are statistically identical. These simulations were run for a simulated model time of 1 Gyr with SWIFT RMVS 3 with the same time step and removal criteria as the leftover planetesimals and E-belt cited above.

The distribution in semi-major axis, eccentricity and inclination for the leftover planetesimals, the E-belt and the A-belt are shown in Figure 1.

### 2.4 Contribution from comets

The cometary impact flux onto the terrestrial planets is caused by the planetesimal-driven migration of the giant planets after the gas of the protosolar nebula has dissipated; the latter occurred at ca. 4563 Ma (e.g. Wang et al., 2016). This relatively 'early' migration (before ca. 4480 Ma; Mojzsis et al. 2019) is caused by the scattering of distant planetesimals beyond Neptune and by mutual scattering amongst the giant planets (Thommes et al., 1999; Tsiganis et al., 2005; Gomes et al., 2005; Morbidelli et al., 2007, 2009, 2010; Brasser et al., 2009; Levison et al., 2011; Nesvorný, 2011, 2015a, 2015b; Nesvorný & Morbidelli, 2012; Brasser and Lee, 2015). The trigger mechanism for this giant planet migration event, however, is unknown and may be stochastic (Levison et al., 2011).

Presently, it is computationally impractical to try to calculate directly the impact probability of comets with the terrestrial planets in one single dynamical N-body simulation. For this reason, we computed the cometary flux onto the terrestrial planets in three sequential steps. In step 1 we determine the best initial conditions for the giant planets and the planetesimal disk that ultimately has the highest probability of reproducing the current architecture of the giant planets. In step 2 we compute the fraction of comets from the trans-Neptunian disk that venture closer than 1.7 au from the Sun. Finally, in step 3 we calculate the impact probability of comets that came closer than 1.7 au from the Sun with the terrestrial planets. Each step is described in detail, next.

### 2.4.1 Initial conditions of the giant planets and planetesimal disk

The evolution of the giant planets during planetesimal-driven migration is chaotic (Tsiganis et al., 2005), and the probability of the planets ending up near their current orbits is low: only a few percent (Nesvorný & Morbidelli, 2012; Brasser & Lee, 2015). Wong et al. (2019) employed a rigorous study of the initial conditions to enhance this probability: they ran test simulations with varying initial conditions and decided which combination of three input parameters best reproduces the current configuration of the giant planets according to the criteria of Brasser & Lee (2015). The free parameters are: the initial semi-major axis of Jupiter, the total mass of the planetesimal disk, and the outer edge of the planetesimal disk. They assumed the planets were initially in a quadruple resonance 3:2, 3:2, 2:1, 3:2. Wong et al. (2019) concluded that the best combination of parameters has Jupiter initially at 5.6 au, a disk mass of 18 $M_E$ and an outer disk edge at 27 au. These parameters are used as input for Step 2.

### 2.4.2 The fraction of comets that enter the inner Solar System

To calculate the impact probability of comets onto the Moon and the terrestrial planets we ran one set of 512 simulations with the symplectic integrator SyMBA (Duncan et al., 1998) for 500 Myr using the initial conditions determined in Step 1; the time step remained the same (0.01 yr). The only difference is that now the comets were removed from the simulation when they ventured closer than 1.7 au, or farther than 1000 au, from the Sun. The vectors and the time when comets cross the 1.7 au barrier are stored for usage below.

Output shows that 22% of all comets from the trans-Neptunian disk crossed the 1.7 au barrier, or the equivalent of about 4 $M_E$. We identified 23 simulations for which the final orbital architecture of the giant planets matches that of the real Solar System. Planetesimals that crossed the 1.7 au barrier of these simulations were combined as initial conditions for the next step.

### 2.4.3 The impact probability of the comets with the terrestrial planets

All comets that we pooled together were subsequently integrated for a model time of 10 Myr using SWIFT RMVS3 with the same time step and removal criteria for the leftover planetesimals and E-belt simulations. These simulations included all eight major planets on their current orbits as well as the principal asteroids Vesta and Ceres. The simulation duration is about two orders of magnitude longer than the average dynamical lifetime of a Jupiter-family comet, which is about 150 kyr (Levison & Duncan, 1997; di Sisto et al., 2009; Brasser & Morbidelli, 2013). The comets were introduced into the simulation as massless test particles with the same position and velocity as when they crossed the 1.7 au barrier from Step 2. We initially added all of the comets to the simulation at once, but to test the validity of this approach we also ran one simulation where the comets were introduced into the simulation at the time they crossed the barrier from Step 2, which ran for 500 Myr. Each simulation contained approximately 650 comets, for a total of nearly 15000 comets.

Both methods of integration – adding all of the comets at once or sequentially – yielded identical results. The comets are dynamically controlled by Jupiter and thus they only spend a few tens of thousands of years in the inner Solar System (Levison & Duncan, 1997; di Sisto et al., 2009; Brasser & Morbidelli, 2013). Their short dynamical survival time, the even shorter time they spend crossing the orbits of the terrestrial planets and the relatively low number of comets meant that we did not record any physical impacts onto the terrestrial planets. We therefore used the Wetherill (1967) averaging procedure to calculate the impact probabilities.

### 3 Monte Carlo impact experiments

In all but the giant planet migration simulations the planetesimals carry no mass. To calculate the amount of mass that impacted the terrestrial planets we need to calculate the number of impacts with the appropriate size-frequency distribution of the planetesimals as it existed at that time. Usually planetesimals are considered to have a power-law size-frequency distribution (SFD) such that the probability of a planetesimal having a diameter between D and D+dD is p(D) ∝ $D^{-\alpha}$. There is no consensus on what SFD to choose for these early impactors, but several works make use of the current SFD of the main asteroid belt (e.g. Bottke et al., 2012; Morbidelli et al., 2012; Morbidelli et al., 2018). Although the main asteroid belt's SFD is wavy today, a simple power-law distribution holds reasonably well for 1≤D≤100 km with α~2 (Bottke et al., 2005; Masiero et al., 2015), but there appears to be a steepening to α~3 for diameters D>100 km. Bottke et al. (2005) show that it may take a few billion years for the wavy SFD to manifest, but this may depend on their initial conditions, so that the timing of the established SFD is inexplicit.

When examining the ratio of small to large lunar craters Minton et al. (2015) proposed an alternative SFD for the inner Solar System planetesimals that differs from that of the main asteroid; they suggested a steepening of the SFD for large objects and that the current SFD of the main asteroid belt is a poor match for early Earth and lunar impactors. This study was followed by Johnson et al. (2016), who analysed some of the Archean terrestrial spherule beds and attempted to constrain the mass of impactors that struck the ancient Earth. Both works suggest that the late accretion impactor SFD had a steep cumulative slope for projectiles with diameters D>50 km. Neither work specifically gives the cumulative slope at the high end, so we have kept it at 3, similar to what is obtained for the largest bodies in the main asteroid belt. Given the uncertainties in the reported observations for the early SFD of these planetesimals it was decided that we adopt the simplest model that we can think of: the cumulative SFD has a power law of α~2.11 for 1≤D≤100 km, and α~3.11 when D>100 km; these values are the best single-power fits for the current asteroid belt's SFD. We change the slope at a diameter of either 50 km (Johnson et al., 2016) or 100 km (Bottke et al., 2005).

When α≤2 the accreted mass is dominated by a few large impactors because the variance is formally undefined (Tremaine & Dones, 1993). For these low values of α, the size-frequency distribution is by definition shallow and it enhances the ratio of the total mass accreted by the Earth versus that of the Moon by a factor of ≥4 over the ratio of their gravitational cross sections (Zahnle & Sleep, 1997; Bottke et al., 2010; Brasser et al., 2016b). For the Earth, we have argued that most of the late accretion came from a single lunar-sized impactor (Brasser et al., 2016b; cf. Sleep, 2016), something which was also implicitly suggested by Bottke et al. (2010); for Mars we have argued that most of its late mass addition was delivered by a single Ceres-sized body (Brasser & Mojzsis, 2017; Woo et al. 2019).

Although not well constrained by observations, the comets are assumed to have a simple SFD that matches the hot Kuiper belt, where the change in the slope occurs at D~140 km and the slopes are 4.35 and 1.8 at the high and low end, respectively (Fraser et al., 2014). We also ran some simulations where the change in the cumulative slope occurs at a diameter of ca. 60 km and cumulative slope indices of 4.8 and 1.9 and the high and low ends, respectively (Fraser & Kavelaars, 2009). Our sensitivity tests did not include a more complicated choice based on the crater size-frequency distribution on Iapetus (Charnoz et al., 2009) with three slopes. We assume that for all planetesimal populations considered here the SFDs are the steady state outcome of collisional evolution, and are thus assumed static with time.

To estimate accurately the amount of mass accreted by each planet, we run Monte Carlo impact experiments of the leftover planetesimals, E-belt planetesimals and the comets onto the planets (Brasser et al., 2016b). In principle the analysis can be performed analytically, but computing the uncertainties is simpler with numerical methods. Figure 2 shows a flow chart of our procedure. For each planetesimal that was generated we compute a set of 5 random numbers uniformly between 0 and 1; each of these is tied to a different planet, e.g. random number #1 is tied to the Moon and #5 to Mercury. The purpose of each of these random numbers is to determine whether or not the planetesimal will collide with a planetary body: if any one of these random numbers is lower than the impact probability with the corresponding planet we assume that the planetesimal strikes the surface of said planet and we generate the next planetesimal. If none of the five random numbers are lower than the impact probabilities with the planets then we assume no impact occurred and we generate the next planetesimal. We sum both the total mass of all the generated planetesimals and how much mass impacts the surface of each planet. The simulation continues to generate new planetesimals until either the Moon has accreted 0.025 wt.% (leftover planetesimals) or if the combined total mass in planetesimals reaches 18 $M_E$ (comets) or $4\times10^{-4}$ $M_E$ (E-belt).

To keep the simulation time reasonable, planetesimals had a minimum diameter of 1 km and an assumed maximum diameter of either 1000 km for leftover planetesimals and the E-belt, and 2000 km for the comets; these large diameters are comparable in size to those of the dwarf planets Eris and Pluto (Brown et al., 2006). We used a bulk density of 1400 kg m$^{-3}$ for the comets, although it should be noted that there may be a density gradient in the cometary population with size, i.e. larger objects could have higher bulk densities than smaller ones. For the asteroidal impacts, we assume a density of 2500 kg m$^{-3}$, which is a value designed to account for all compositions and the assumed impactor's SFD (Morbidelli et al., 2012). We did not include Vesta and Ceres in the Monte Carlo code because we prefer to focus our attention on the terrestrial planets.

Following Brasser et al. (2016b) we further keep track of how many basins (defined as craters with diameter >300 km) as well as craters with diameter >20 km are formed on the terrestrial planets. The diameter of the final craters depends on the impact velocity, impact angle and material properties of impactors and target planet. The relationship between the final diameter of a crater excavated on the surface of a target body and the diameter of the impactor that creates it is complicated (Melosh, 1996). Upon impact, a transient crater with diameter $D_{tr}$ is created, whose diameter scales with impactor diameter, $D_i$, as (Schmidt & Housen, 1987)

$$D_{tr} = 1.16 \left(\frac{\rho_i}{\rho_t}\right)^{1/3} D_i^{0.78} v_i^{0.43} g_t^{-0.22} \quad (1),$$

where subscript $i$ stands for the impactor and $t$ for the target. Here $\rho$ is the bulk density, $v$ is the impact speed and $g$ is the acceleration due to gravity. The units are mks. For small, simple craters the final crater diameter, $D_f$, is related to the transient crater diameter $D_{tr}$ as $D_f \sim 1.25 D_{tr}$ (Chapman & McKinnon, 1986), while for larger, more complex craters a transition occurs that relaxes the outer walls, which thus further increases its diameter. For these larger craters the final diameter can be computed via (Croft, 1985)

$$D_f = \frac{D_{tr}^{1+\eta}}{D_{SC}^{\eta}} \quad (2),$$

where η=0.15-0.18 and where $D_{SC}$ is the crater diameter at which the transition from simple to complex craters occurs. Setting η=0.15 we finally have that the final crater diameter, $D_f$, scales as the impactor diameter $D_i^{0.90}$. The value of $D_{SC}$ for the Moon is approximately 15 km (Pike, 1980) while for the other silicate worlds it can be calculated from (Pike, 1980)

$$D_{SC,p} = \frac{g_{Moon}}{g} D_{SC,M} \quad (3).$$

Assuming an impact angle θ=45°, an impactor speed $v_i$=16 km s$^{-1}$ and impactor density $\rho_i$=2500 kg m$^{-3}$ on the Moon the diameter of a planetesimal excavating a 20-km crater should be $D_i$=1.1 km and $D_i$=22 km for producing a basin. Similarly, the number of craters and basins on Mercury and Mars are calculated in this fashion although we substitute the average impact speeds with these planets obtained from the dynamical simulations.

## 3.1 Impact erosion of planetary atmospheres

Impacts may be capable of eroding some of a planet's atmosphere (Melosh & Vickery, 1989). Here we follow Schlichting et al. (2015) to compute the expected amount of atmospheric mass loss from our late accretion mass production functions.

Mass losses of the atmosphere due to impacts occurs in two regimes. For planetesimals with mass $m_{imp} > \frac{1}{2}\rho_a(\pi h D)^{3/2}$ all of the atmosphere above the tangent plane of the impact site will be ejected. Here $\rho_a$ is the atmospheric surface density, $h$ is the scale height and $D$ is the impactor diameter. The total ejected atmospheric mass from such impacts amounts to a fraction of $\sim \frac{1}{4}\frac{h}{D}$ of the total atmospheric mass. Smaller impactors, but with mass $m_{imp} > 4\pi\rho_a h^3$, however, are only able to eject a small fraction of the atmospheric mass above the tangent plane. Combining the two regimes, the amount of atmospheric mass loss per impact is given by

$$m_{ej} = m_{imp} \, min\left[\frac{r_{min}}{D}\left(1 - 4\left(\frac{r_{min}}{D}\right)^2\right), 8\mu_{cap}\left(\frac{r_{cap}}{D}\right)^3\right], \quad \text{where} \quad \mu_{cap} = \left(\frac{2h}{\pi R_p}\right)^{1/2}, \quad r_{cap} = (2\pi)^{1/6}\left(\frac{3\rho_a}{4\rho_i}\right)^{1/3}(hR_p)^{1/2} \text{ and } r_{min} = h\left(\frac{3\rho_a}{\rho_i}\right)^{1/3}.$$ No atmospheric escape occurs for $D<2r_{min}$; when $D>2r_{cap}$ the whole atmosphere above the tangent plane is ejected.

We track the amount of atmospheric loss from such impacts on Venus, Earth and Mars in our Monte Carlo simulations to estimate how severe early atmospheric erosion due to impacts was on these planets. Since the composition and surface pressure of the primordial atmospheres of these planets are unknown we assumed the following parameters: $h=10$ km for all three bodies (an intermediate value of the scale heights of their present atmospheres), while $\rho_a=5$ kg m$^{-3}$ for Venus and Earth, and $\rho_a=1$ kg m$^{-3}$ for Mars, respectively. These values correspond to Hadean, (early) pre-Fortunian and early Noachian surface pressures of 5 bar on Venus and Earth, and 0.38 bar on Mars, respectively. The total atmospheric mass is $m_a = 4\pi\rho_a R_p^2 h$. With our chosen parameters $r_{min}=1.96$ km for Venus and Earth and 1.1 km for Mars, and $r_{cap}=42$ km for Venus and Earth and 18 km for Mars. Due to the uncertainties in the initial atmospheric pressure and composition we did not perform sensitivity tests of these parameters and leave that for future work.

**4 Outcomes**

**4.1 Impact probability, impact velocity and total accreted mass**

We plot the average impact probabilities, impact velocities and mass augmentations accreted by the planets for all three source populations in Figure 3.

The data are listed in Tables A1 to A4 in the Appendix of this paper (see also Mojzsis et al., 2019). All uncertainties are reported as 2σ and are almost entirely the result of rare and stochastic impacts with high mass in different initialisations of the Monte Carlo simulations (Bottke et al., 2010). Uncertainties in the impact probabilities are far smaller than the uncertainties caused by stochastic accretion of a few large objects. For Vesta and Ceres the listed values of mass augmentation are estimates obtained from the dynamical simulations only. For the leftover planetesimals and the E-belt contribution the accreted mass values in the last column are for the simulations where the changeover radius is 100 km. For the comets we settled for a changeover radius of 60 km and the size-frequency distribution of Fraser et al. (2014).

No actual impacts occurred in the dynamical simulations of the comets so that the plotted uncertainties in the impact probabilities and velocities only come from intra-simulation variations, and are certainly underestimates. Furthermore, we assumed that upon impact all the mass that strikes the inner planets, Moon, Vesta and Ceres is accreted with 100% efficiency. In practise this is likely to be less (e.g. Raymond et al., 2013; Kraus et al., 2015), so that the amount of mass that we list here striking the surfaces is regarded as a *maximum* amount that is actually retained. We caution that the typical impact velocity increases with time as the planetesimals become dynamically more excited due to secular effects and repeated encounters with the planets; the values listed here, however, are averaged over 0.5 Gyr (leftover planetesimals) or 1 Gyr (E-belt). For the E-belt our impact probabilities for the Earth, Moon and Mars agree with those reported in Bottke et al. (2012); the values listed for leftover planetesimals and comets are new. Of the leftover planetesimals 5.1% are ejected from the Solar System and 53.4% strike the Sun because of the influence of the $\nu_6$ secular resonance (Morbidelli et al., 1994); for the E-belt population 8.1% is ejected and 70.5% collides with the Sun.

It is clear that the amount of mass striking the rocky bodies in the inner Solar System varies markedly (see bottom panel of Figure 3). First, Ceres and Vesta are pummelled by a very large amount of material, especially from comets, which in total approaches 0.5–1 wt.%. The amount of mass colliding with Vesta from the E-belt is substantial, and similar to that coming from the main belt. If all of this came in the form of a single planetesimal with a density of 2500 kg m$^{-3}$ then its diameter would have been 66 km. Assuming that this planetesimal struck Vesta at 9 km s$^{-1}$ it would excavate a crater with a diameter of approximately 550 km using the crater scaling laws employed here, comparable in size to the Rheasilvia basin (D~450 km) near Vesta's south pole (Marchi et al., 2012a; Schenk et al., 2012).

For most planets, however, the mean mass augmentation reported is low. Brasser et al. (2016b) showed that the mass augmentation on Earth as deduced from its HSE abundance in the mantle is at >99.9 percentile of the cumulative distribution when the SFD had no steepening at the high end. In contrast, for Ceres and Vesta the 50$^{th}$ percentile approaches a 0.5 wt.% increase.

Mars, Mercury and the Moon fall into the second category: these three bodies are also struck by a large amount of material, and a high fraction of comets for the Moon and Mars. Despite Mercury's proximity to the Sun and its smaller surface area than Mars it experiences a higher probability of impacts from leftover planetesimals. We think that unlike Mars, Mercury has great difficulty in scattering material in its proximity onto an orbit that removes it from its sphere of influence altogether so that more material that crosses its orbit is destined to impact the planet.

Earth and Venus form the third category because they proportionally accrete much less than other inner solar system bodies. In absolute terms the Earth and Mars accrete the most, although in the absence of direct samples it is not known what the absolute late accretion amount was on Venus and Mercury. In Figure 4 we show the differential mass augmentation profile of late accretion to the Earth, Moon and Mars from all three sources. To create the plot we had to assume that all the impactors had the same mass, but the total accreted mass was taken from Tables A1-A3, which takes the effect of the SFD into account. As such, the much larger fraction of leftover mass accreted by the Earth than the other planets comes from the Earth accreting proportionally more large objects. The accretion profile assumes the comets began arriving just after the Moon's formation at 4.5 Ga (Mojzsis et al., 2019), although we acknowledge that the comets could have arrived earlier. There are three distinct peaks for each population, with the cometary peak arriving before the leftover and E-belt peaks.

We compute the background mass in small leftover planetesimals at 4.5 Ga to be $9.9^{+7.1}_{-6.9} \times 10^{-4}$ $M_E$. This mass is computed by dividing the amount of mass striking the Moon by its impact probability. The HSE abundances in the terrestrial and martian mantles indicate that they both accreted a further 0.7 wt.% and 0.8 wt.% respectively (Day et al., 2016). We can perform the same exercise using the Earth as the reference rather than the Moon. Doing so yields a leftover planetesimal mass of 0.05 $M_E$, while scaling it to Mars the leftover planetesimal mass is increased to 0.1 $M_E$. These factor of ~2 difference in the calculated leftover planetesimal masses are expected because the Moon does not appear to have been struck by any planetesimal with a diameter greater than about 170 km (Potter et al., 2012). When anchoring our results to the Moon the mass in leftover planetesimals is very low and potentially violates constraints on the planetesimal mass from dynamical models of terrestrial planet formation (Jacobson & Morbidelli, 2014; Brasser et al., 2016a).

An independent estimate of the total mass in leftover planetesimals is provided from the Hadean zircon record and therefore, of limits to the time in which Earth could have experienced wholesale crustal melting. Abramov et al. (2019) claim that if Earth accreted in excess of $9.3 \times 10^{21}$ kg by late accretion in the time subsequent to giant planet migration the Hadean crust would have effectively been entirely molten and its zircon record would be completely reset. Since it appears from Figure 3 and Tables A1 to A4 that roughly 3/4 of the background accreted mass on Earth came from leftover planetesimals, then the total mass in such planetesimals at 4.5 Ga is $3/4 \times 9.3 \times 10^{21}/0.13 \sim 5.3 \times 10^{22}$ kg or 0.01 $M_E$. Based on the above arguments, it is evident that the crust was destroyed at that time (Mojzsis et al, 2019). The $9.3 \times 10^{21}$ kg value had to have been accreted onto the Earth before 4.38 Ga, which is the age of the oldest known terrestrial zircon (Valley et al. 2014). The expected amount of mass that would have struck the Moon is then $5.3 \times 10^{19}$ kg because the impact probability with the Moon is just 0.9%, corresponding to about 0.07 wt.%, i.e. three times as much as suggested from the lunar HSE budget (Day & Walker, 2015), and about 50% more than computed from tungsten isotopes (Touboul et al., 2015).

Morbidelli et al. (2018) dismissed the lunar HSE constraints because it has been suggested that the lunar HSE abundance is not indicative of accretion going as far back as 4.5 Ga. Hence the lunar HSE budget only reflects addition after a particular epoch; Zhu et al. (2019) advocates that the lunar HSEs are only retained after 4.35 Ga. In our study we argue instead that the lunar HSEs pose a powerful constraint going all the way back to lunar formation and that they, the lunar tungsten isotopes and the Hadean zircons, are the best way to constrain the mass in leftover planetesimals.

In summary, we expect the total mass in leftover planetesimals to have been in between 0.001 $M_E$ and 0.01 $M_E$, which does not include the large singular late veneer-scale impactors that struck the Earth (Brasser et al., 2016) and Mars (Brasser & Mojzsis, 2017) and delivered their HSEs. Nor does this result consider mass augmentation to Earth from the earlier Moon-forming event.

The impact probabilities for the comets should be interpreted with care because the values reported here assume that the comets are treated as asteroids and do not undergo any physical evolution. We know from observations that comets lose mass and fade, so this assumption is probably not correct. The aspect of cometary fading, sublimation and disruption has been studied at length (e.g. di Sisto et al., 2009; Fernandez et al., 1999; Levison & Duncan, 1997; Levison et al., 2002; Brasser & Morbidelli, 2013; Brasser & Wang, 2015; Rickman et al., 2017). Recently, Nesvorný et al. (2017b) concluded from observations and a high number of dynamical simulations of the evolution of comets that the number of revolutions with perihelion distance q<2.5 au that a typical short-period comet is able to survive is $N_p(2.5) \sim 500(D/1 \text{ km})^\beta$ for comet diameters D≤100 km, and is approximately constant for D>100 km. The value of β depends on the mechanism of mass loss. If comets have a fixed mass loss rate per orbit, which some observations appear to indicate (Reach et al., 2007), then we have β=1/3. If instead the radius shrinks by a fixed amount per orbit (Fernandez, 2010), then β=1. It is not clear which mechanism dominates, and thus we adopt β=1/2 to compute the crater densities on the Moon, Mars and Mercury from the comets. Most of the mass is carried by the large comets and thus the total delivered mass is only expected to be somewhat smaller than listed in Table A3 due to the sublimation. However, the sublimation effect is important for the chronology functions because these are pegged to relatively small crater diameters.

**4.2 Lunar, martian and mercurian impact chronologies**

**4.2.1 The Moon**

In our Monte Carlo impact experiments we follow Brasser et al. (2016b) and use an impactor diameter of 22 km as threshold for creating a lunar basin (defined as a crater with diameter D>300 km). We further keep track of the number density of lunar craters we produce with diameter 20 km or greater. The density of these craters is used to derive the chronology function. We use this crater diameter rather than the standard 1 km diameter craters because the former are thought to be created by projectiles with D=1 km. The 1-km in diameter craters are formed by impacts through very small planetesimals that probably are created by inter-planetesimal collisions; these small bodies further experience non-gravitational forces changing the relative total numbers at different sizes. At present we cannot model both of these effects in the dynamical simulations, so that we settle for larger crater diameters. The ensuing crater density, $N_{20}$, is computed by counting the number of lunar impacts in the Monte Carlo simulations and dividing by the lunar surface area; it does not take into account apex-antapex asymmetries or other variations in the crater density.

If the largest object that struck the Moon had a diameter of 170 km then the expected number of craters with diameter >20 km is approximately $(170/100)^3 \times 100^2 = 49130$. From the Monte Carlo simulations, we find that from 4.5 Ga onwards, there are $46414^{+30746}_{-31755}$ (2σ) such impacts on the Moon from leftover planetesimals, a further $9004^{+2623}_{-3199}$ from the E-belt and $304^{+96}_{-111}$ from the main belt; the comets contribute a further $3739^{+24}_{-28}$ craters, and we have taken the sublimation of the comets into account. The uncertainties listed for the number of cometary craters are those from multiple iterations of the Monte Carlo simulations caused by the adopted size-frequency distribution. In reality we expect the uncertainty to be much larger due to the uncertainty in the sublimation mechanism of the comets. In any case, at 4.5 Ga we have theoretically $N_{20}=1.56^{+0.85}_{-0.90}\times10^{-3}$ km$^{-2}$ from non-cometary sources.

Traditional crater chronology curves are given in terms of $N_1$, i.e. the number density of craters with diameter greater or equal to 1 km, rather than $N_{20}$. We can either proceed to scale up our results to $N_1$ or downscale the traditional crater chronology curves to $N_{20}$. We opt for the latter.

The traditional calibrated lunar chronology is based on the lunar production function of Neukum (1983), who computes $N_1/N_{20}=995$. For Mars Ivanov (2001) and Werner and Ivanov (2015) compute $N_1/N_{20} = 419$; this value was also used by Werner et al. (2014). Note that these two scaling factors are both much lower than the value of 1400 used by Morbidelli et al. (2012). The chronologies of Neukum and Werner report a value of $N_{20}$ of $4.54\times10^{-3}$ km$^{-2}$ and $1.00\times10^{-3}$ km$^{-2}$ at 4.5 Ga respectively.

The modelled shape of the crater chronology curve depends in part on the underlying dynamics of the planetesimals, in particular how the number of impacts on each planet varies with time. In an earlier study (Brasser et al., 2016b) we used the decline in the total number of planetesimals as a proxy for the lunar crater chronology. This total decline, however, is not the same as what is recorded on the surfaces of the individual terrestrial planets because the total decline of the population is a combination of several effects: impacts onto the planets, impacts onto the Sun and ejection from the Solar System. Each of these three processes occurs at their own pace and the combination of these planetesimal sinks results in the approximately stretched exponential decrease that was published earlier. Yet, the surfaces of the planets only record one sink: collision with that planet. As such, we need to base the chronology of each planet on the actual impacts recorded on its surface in the simulation.

Thus, we have taken care to draw the chronology curves from actual recorded impacts with each planet in the dynamical simulations. While the Moon was not included in these simulations, we can use impacts on the Earth as a proxy for the lunar chronology curve. In Figure 5 we plot the cumulative distribution of the individual impacts onto the Earth as a proxy for the Moon from the leftover planetesimals (blue) and the E-belt (green) and comets (cyan). All the impacts that are plotted here are from combining the output from the simulations. The curves are normalised to the values of $N_1$ computed at 4.5 Ga in the previous paragraph. We also plot the best fits through these data as dashed lines together with the cumulative chronologies of Neukum et al. (2001) in orange and Werner et al. (2014) in purple. Both the Neukum and Werner curves have an additional linear component caused by a supposed contribution from the asteroid belt and near-Earth object (NEO) population. This component, which is thought to shallow the curves from about 3.7 Ga onwards, has been included here for clarity in our 'total profile' curves. We did not include the Robbins chronology (Robbins, 2014) because its e-folding time for ages older than ca. 4 Ga is only 4.5 Myr which is inconsistent with the dynamical timescales. The chronology of Marchi et al. (2009) has an e-folding time of only 127 Myr and was not included either because it would deviate even further from the N-body data than the Neukum chronology curve. The solid calibration points are from Neukum (1983). We have created and added a few other calibration points as follows.

Current estimates for the age of Imbrium basin hover around 3.95 Ga based on interpretations of U-Pb and $^{40-39}$Ar ages of lunar rocks (e.g. Tera et al. 1974; Schäffer & Husain, 1974; Maurer et al., 1978; Mojzsis & Ryder, 2001; Shearer & Borg 2006). Crater counts near Imbrium basin performed by Head et al. (2010) show that $N_{20}=0.032\times10^{-3}$ km$^{-2}$. Apollo 16 and 17 samples reveal a suite of ca. 4.2 Ga ages in $^{40-39}$Ar (Schäffer & Husain, 1974; Fernandes et al., 2013) and Re-Os (Fischer-Gödde & Becker, 2012) which, given the landing sites' proximity to Nectaris basin, could imply that the age of Nectaris is 4.2 Ga. Further, U-Pb zircon dating of Apollo 14 materials (Norman & Nemchin, 2014; Hopkins & Mojzsis, 2015) yield abundant ca. 4.2-4.25 Ga ages that may be re-distributed ages from basin-forming impacts at that time (Kelly et al. 2018).

Is must be emphasized that this age interpretation for Nectaris is fraught with problems, and that there is no consensus on the time of its formation (e,.g. Reimold et al., 1985). Different authors assign different ages to Nectaris, which can be as young as ca. 4.0 Ga (e.g. Maurer et al., 1978; Schäffer & Husain, 1973). Norman et al. (2016) find a suite of ages in lunar breccia 67955 that cluster near 4.2 Ga. They argue that this sample crystallized in a central melt sheet and that the impact that formed it occurred in the Procellarum-KREEP Terrain rather than in the feldspathic highlands. These petrological and geochemical characteristics imply transport of the clast to the Apollo 16 site as Imbrium ejecta rather than Nectaris ejecta, casting a 4.2 Ga age of Nectaris in doubt. Given the uncertainties we have taken the bold step to proceed to assume a 4.2 Ga age for Nectaris as a calibration point, but have also tested the chronology without it.

Head et al. (2010) reported that at Nectaris basin $N_{20}=0.131\times10^{-3}$ km$^{-2}$. The last data point takes the highest crater counts found by Head et al. (2010) of $N_{20}=0.280\times10^{-3}$ km$^{-2}$ in the lunar highlands, which we take to be a lower estimate of the crater density on the oldest lunar crust, whose age likely corresponds to the oldest known lunar zircon of 4.42 Ga (Nemchin et al., 2009). These three proposed calibration points are shown as open circles. Uncertainties have been set to 0.07 Ga.

Inspecting Figure 5 it is clear that the slope of the flux of leftover planetesimals onto the Moon changes with time, becoming ever shallower. This is entirely expected: planetesimals in the vicinity of Venus and the Earth are quickly accreted or ejected, on a timescale of ~20 Myr (Wetherill, 1977; Chyba, 1990). This is the reason for the initially steep slope between 4.5 Ga and 4.3 Ga. Once most of the planetesimals near Venus and Earth have been eliminated, the Earth-Moon system will continue to be bombarded by planetesimals that are leaking in from the inner asteroid belt region on Mars-crossing orbits. Encounters with Mars are inefficient, and the timescale for Mars to clear planetesimals in its vicinity is of the order of 100 Myr (Michel et al., 2000). This slow delivery of planetesimals from Mars and beyond dominates the bombardment rate from 4.3 Ga to about 3.7 Ga. After 3.7 Ga the bombardment rate appears to drop sharply. This is a result of small-number statistics: by this time, the average number of planetesimals in each simulation of leftover planetesimals is just 42, which is <4% of the starting value, and subsequent impacts are rare. Thus the fits are only reliable on the interval 4.5-3.7 Ga. The E-belt simulations begin to suffer from this artificial slowdown effect near 3.6 Ga.

One may think that this artefact caused by small-number statistics can be prevented by running more simulations. While this is partially true, the better solution relies on starting each simulation with more (massless) planetesimals. It is well known that several planetesimals are able to find meta-stable niches where they are protected against close encounters with nearby planets on long timescales (Duncan & Levison, 1997; Malyshkin & Tremaine, 1999). The Hungaria asteroid family near 1.8 au presents such an example (Bottke et al., 2012). As the simulation continues, these meta-stable planetesimals remain while the rest are eliminated. Eventually there will be a point in time where these meta-stable planetesimals dominate the population and the impact rate will drop substantially because there are no more planetesimals on planet-crossing orbits. To mitigate this effect, the simulations should be started with more planetesimals so that any that survive for longer but are still unstable continue to strike the planets; however, this comes at the expense of longer simulation time and more planetesimals winding up in the meta-stable niches.

From Figure 5 it is clear that both the shape of the chronology curves for all populations as well as their sum are different from that predicted by Werner et al. (2014) and Neukum et al. (2001). The rate of delivery of material from the E-belt is similar to both chronology curves, with an e-folding time in between the Werner and Neukum chronologies, because the rate is dominated by interaction with Mars; the absolute number of impacts is, nonetheless, lower.

**4.2.2 Mars**

The same procedure from the lunar chronology can be applied to Mars. We record $27166^{+17945}_{-18487}$ impacts from leftover planetesimals on Mars that excavate 20-km craters or greater, a further $23074^{+6773}_{-8159}$ from the E-belt and $1573^{+464}_{-573}$ from the main belt. In Mars' case the threshold planetesimal diameter is D~1.6 km owing to the lower impact speed on Mars than onto the Moon, and Mars' higher acceleration due to gravity. The comets contribute an additional $20496^{+77}_{-62}$ craters with diameter of 20 km or greater. Thus on Mars we compute a theoretical $N_{20}=0.502^{+0.17}_{-0.17}\times10^{-3}$ km$^{-2}$ at 4.5 Ga only from inner Solar System objects.

The calculated martian chronology is shown in Figure 6, where the Neukum curve has been replaced with that of Ivanov (2001). The Ivanov and Werner chronologies give $N_{20}$ values of $2.24\times10^{-3}$ km$^{-2}$ and $0.50\times10^{-3}$ km$^{-2}$ at 4.5 Ga respectively.

From ca. 4.4 Ga onwards our analysis shows that the bombardment on Mars is dominated by the E-belt, it may also be true that this is the case from 4.45 Ga onwards if the comets play a less significant role than shown here. We see a similar, but less pronounced, behaviour to the lunar chronology in the shape of the deep blue curve from leftover material: there is a steeper component from planetesimals that are eliminated quickly, and then a slowing of the bombardment rate from approximately 4.3 Ga onwards. In the dynamical simulations Mars experiences 80% fewer impacts than the Earth so that the martian chronology begins to suffer from small-number statistics at 3.8 Ga onwards for both populations. From 4.3 Ga to 3.8 Ga the slope of the combined impacts from leftover planetesimals and the E-belt is somewhat shallower than that proposed by Werner et al. (2014), and is wholly inconsistent with that of Ivanov (2001).

### 4.2.3 Mercury

Our Monte Carlo simulations yield $363342^{+240500}_{-247220}$ impacts on Mercury from leftover planetesimals that create 20-km craters, an additional $33229^{+9710}_{-11802}$ impacts from the E-belt and $1094^{+322}_{-398}$ from the main belt; comets potentially contribute an additional $2305^{+18}_{-23}$ impacts. This leads to very high crater densities on Mercury, $5.32\times10^{-3}$ km$^{-2}$ from non-cometary sources, with the leftover population contributing the far majority and thus dominating the mercurian impact bombardment before 4.2 Ga. The (theoretical) chronology for Mercury is provided in Figure 7. The Werner chronology for Mercury overlaps well with the slope of our E-belt contribution, but begins to deviate for ages older than 4.3 Ga because of the high flux of leftover planetesimals. The Ivanov and Werner chronologies give $N_{20}$ values of $5.21\times10^{-3}$ km$^{-2}$ and $1.78\times10^{-3}$ km$^{-2}$ at 4.5 Ga respectively.

We list the total crater densities from each population and their combined total at 4.5 Ga in Table 1. The fading factor for the comets has been taken into account.

| Body | $N_{20}$ Leftover [$10^{-3}$ km$^{-2}$] | $N_{20}$ E-belt [$10^{-3}$ km$^{-2}$] | $N_{20}$ A-belt [$10^{-3}$ km$^{-2}$] | $N_{20}$ Comets [$10^{-3}$ km$^{-2}$] | $N_{20}$ Total [$10^{-3}$ km$^{-2}$] |
|---|---|---|---|---|---|
| Moon | $1.22^{+0.81}_{-0.84}$ | $0.237^{+0.06}_{-0.08}$ | $0.008^{+0.002}_{-0.003}$ | $0.098^{+0.001}_{-0.001}$ | 1.564 |
| Mars | $0.189^{+0.108}_{-0.141}$ | $0.160^{+0.041}_{-0.071}$ | $0.011^{+0.003}_{-0.004}$ | $0.142^{+0.0004}_{-0.0008}$ | 0.502 |
| Mercury | $4.86^{+3.21}_{-3.31}$ | $0.444^{+0.130}_{-0.159}$ | $0.015^{+0.004}_{-0.005}$ | $0.031^{+0.0004}_{-0.0005}$ | 5.350 |

Table 1: Crater densities $N_{20}$ from the Monte Carlo simulations on the Moon, Mars and Mercury from all four small body sources, and their combined total density at 4.5 Ga.

### 4.2.4 Chronology function

In most cases the decline of each population shown in Figures 5 to 7 is not a simple exponential, which in the log-linear coordinates of the figures should show up as a straight line. Instead the slope decreases with time so that we fit the dynamical impact chronology from inner solar system objects as a superposition of several exponentials of the form

$$N_{20} = \sum_{j=1}^{3} n_j \sum_{i=1}^{N} a_i \, e^{\frac{t-4500}{\tau_i}} \quad (4),$$

where $t$ is time from the present in Myr, $a_i$ and $\tau_i$ are the fitting constants and e-folding times with as many terms as needed, and $n_j$ are the values of $N_{20}$ at 4.5 Ga from each non-cometary reservoir. The fitting parameters are given in Table 2; the values of $n_j$ are listed in columns two to four of Table 1.

Empty columns imply that additional terms did not improve the fit. The main belt fits to Mars and Mercury are rather uncertain because in our dynamical simulations of this reservoir very few impacts were recorded.

| Moon (Earth) leftover | | | |
|---|---|---|---|
| $a$ | 0.04298 | 0.5766 | 0.3779 |
| $\tau$ [Myr] | 6.208 | 27.07 | 79.33 |
| Moon (Earth) E-belt | | | |
| $a$ | 1.000 | | |
| $\tau$ [Myr] | 175.33 | | |
| Moon (Earth) A-belt | | | |
| $a$ | 0.5660 | 0.3703 | |
| $\tau$ [Myr] | 64.69 | 247.39 | |
| Mars leftover | | | |
| $a$ | 0.1666 | 0.8683 | |
| $\tau$ [Myr] | 8.062 | 66.11 | |
| Mars E-belt | | | |
| $a$ | 0.4369 | 0.5753 | |
| $\tau$ [Myr] | 79.51 | 227.37 | |
| Mars A-belt | | | |
| $a$ | 0.2774 | 0.7092 | |
| $\tau$ [Myr] | 36.44 | 338.92 | |
| Mercury leftover | | | |
| $a$ | 0.8385 | 0.1581 | |
| $\tau$ [Myr] | 44.72 | 109.84 | |
| Mercury E-belt | | | |
| $a$ | 1.1097 | | |
| $\tau$ [Myr] | 205.25 | | |
| Mercury A-belt | | | |
| $a$ | 0.9449 | | |
| $\tau$ [Myr] | 177.07 | | |

Table 2: Fitting constants ($a$) and e-folding times ($\tau$) for the dynamical chronology functions from leftover planetesimals, the E-belt and the A-belt on the Moon, Mars and Mercury.

The same functional form can be applied to the mass decline of the comets as they rush through the inner Solar System. Accretion profiles for each planet are not available since we did not record any physical impacts. The temporal mass of comets passing through 1.7 au is then of the form

$$M_c = 4 M_E \sum_{i=1}^{N} a_i\, e^{\frac{t-4500}{\tau_i}} \quad (5),$$

where the fitting constants are listed in Table 3.

| Mass profile of cometary material | | | |
|---|---|---|---|
| $a$ | 0.1511 | 0.6782 | 0.2340 |
| $\tau$ [Myr] | 0.919 | 6.440 | 41.82 |

Table 3: Fitting constants and e-folding times for the cometary decline.

From Tables 2 and 3 we conclude that by 4.38 Ga about 91% of the mass in leftover planetesimals, 50% of the E-belt and 98% of the comets that will strike the Earth have already done so. The oldest martian and lunar zircons have ages between 4.47 Ga and 4.42 Ga (Bouvier et al., 2018; Nemchin et al., 2009) which indicates that the crusts of these worlds have not been entirely molten since that time. By ca. 4.47 Ga, we calculate that 38% of leftover planetesimals, 16% of the E-belt and 85% of the comets that will collide with Mars have already done so. For the Moon these values are 85% of leftover planetesimals, 37% of the E-belt, and 97% of the comets. Abramov & Mojzsis (2016) have shown that it is more difficult to melt Mars' crust than Earth's, so that even though by 4.47 Ga Mars will still accrete substantial amounts of mass, it will not be sufficient to cause wholesale crustal melting.

From about 4.3 Ga onwards the time profile of the bombardment on the Moon from leftover planetesimals and the E-belt has an e-folding time of approximately 180 Myr because by this time impacts are dominated by the contribution from the E-belt. On Mars and Mercury the situation is the same. This supports our earlier claim that on long timescales the bombardment on the terrestrial planets is dominated by material that is leaking in from the asteroid belt (E-belt) on Mars-crossing orbits (Wetherill, 1977). Our result contradicts that of Morbidelli et al. (2018), who suggested a much higher contribution from leftover planetesimals. In their simulations said population has a much slower rate of decline than in ours due to their different initial conditions. Even though Morbidelli et al. (2018) also used initial conditions arriving from planet formation simulations, many of the planetesimals therein were clones while ours are all unique.

The fits to the lunar and martian chronology functions have somewhat different constants and e-folding times, yet it is unclear from the figure if the two distributions are statistically different. Performing a Kolmogorov-Smirnov test on the N-body impact data reveals that for leftover planetesimals the probability of agreement between the two distributions is 0.3% while for the E-belt the agreement is just 1.25%. Hence, the two chronology curves are statistically different and we argue that simple extrapolation of the lunar chronology curve and applying it to Mars should be done either with extreme caution or be avoided altogether. This disagreement is not the result of the small-number impacts from approximately 3.7 Ga onwards but is due to the slightly different e-folding times.

**4.2.5 Basin formation and the origin of the 'classical' late heavy bombardment (ca. 3.95 Ga)**

Our Monte Carlo simulations yield $82^{+54}_{-57}$ lunar basins from leftover planetesimals. The E-belt delivers a further $16^{+8}_{-8}$, which is higher than that suggested by Bottke et al. (2012), and the main belt another $1^{+1}_{-1}$. The difference can most likely be ascribed to our usage of a different size-frequency distribution of the impactors. Minton et al. (2015) have suggested that the number of lunar basins calls for a steep decline with increasing diameter in the number of large lunar impactors, with a possibly steeper cumulative slope than we employed here. Miljkovic et al. (2013) also indicated that impactors striking the nearside will produce a larger basin than on the farside, complicating matters considerably. The total number of calculated lunar basins is much higher than the inferred 42 basins from observations by the GRAIL mission (Neumann et al., 2015) or listed by Wilhelms (1987). The number of calculated lunar basins is consistent with observations within 2σ, although the comets could contribute a further ~100 basins. Given the rapid decline in the number of basins generated per unit time from both the comets and leftover planetesimals we argue that most of the basins were created in the first 80 Myr, between 4.5 and 4.42 Ga. Thus most basins formed before the age of the oldest known lunar zircon of 4417 Ma (Nemchin et al., 2009); this age is attributed to lunar magma ocean solidification and crust formation. We postulate that the lower number of observed basins on the lunar surface compared to the Monte Carlo simulation output may be the result of the earliest basins not having survived in a warmer and more pliable 'taffy-like crust' during the lunar magma ocean phase.

The martian surface has been substantially modified since its formation, so that basin erasure and burial is expected. The expected number of basins from leftover planetesimals and the E-belt are $49^{+35}_{-34}$ and $41^{+17}_{-17}$, respectively, with the main belt creating another $1^{+1}_{-1}$. This is about a factor of three higher than the ~30 observed basins on Mars listed in Werner (2014) or the 17 listed in Werner (2008). Bottke and Andrews-Hanna (2017) instead focus on crustal-scale impact basins with diameter $D_f>780$ km, all of which excavated a significant fraction (>25%) of the crust, and they suggest that three clearly visible major basins (Hellas, Isidis, and Argyre) and the buried Utopia basin fit this criterion. The number of martian basins remains disputed, however. Frey (2008) identifies many more candidate basins based on subtle topography and gravity signatures and suggests (controversially) that the total average density of martian basins from visible and buried basins is $N_{300}=2\times10^{-6}$ km$^{-2}$, which when multiplied by Mars' surface area results in 288 basins; this is much higher than that inferred from the simulations. The comets contribute on average a further 890 basins on Mars, which is also well beyond the observed and inferred values cited above.

We again attribute the differences between the calculated and observed numbers to most basins having been created when the martian crust was under formation, or before the potential crustal reset at 4.48 Ga as suggested by Bouvier et al. (2018). This requires that the comets arrived before 4.48 Ga (Mojzsis et al., 2019). Bottke & Andrews-Hanna (2017) suggest that fewer than 12 basins with $D_f$>780 km formed between 4.5 Ga and 4.1 Ga, and that the four giant martian basins Hellas, Isidis, Argyre and Utopia all formed between 3.8 Ga and 4.1 Ga. Ignoring for now the age constraints on these basins, our model predicts the formation of ~10-15 such basins with $D_f$>780 km after 4.5 Ga. This is a factor of a few higher than what is observed on the crust of Mars today, but is far below the 20 basins with $D_f$>1000 km suggested by Frey (2008) and consistent with the ≤16 proposed by Bottke & Andrews-Hanna (2017). This number discrepancy is not caused by our choice of size-frequency distribution: we calculate that a crater with $D_f$>780 requires an impactor with diameter $D_i$>75 km. In the current main asteroid belt the ratio between the number of objects creating a martian crater with $D_f$>20 km versus those that excavate a crater with $D_f$>780 is ~2000, while from our simple power law SFD we get a ratio of 3300. When we assume that the change in the SFD slope occurs at 50 km diameter, the ratio is increased to ~5000. Extrapolating to these large basins from our value of $N_{20}$ we predict fewer basins with $D_f$>780 km than using the main belt SFD. We are uncertain as to what causes the disagreement between the predicted and observed number of large basins on Mars, although small-number statistics may play a role. The chronology proposed by Bottke & Andrews-Hanna (2017) is not entirely consistent with ours: assuming that all these large basins come from the E-belt (which is the source with the slowest rate of decline), the expected timing of the latest of these large basins is at 4.0 Ga and the last four formed between 4.3 Ga and 4.0 Ga; it is more likely that some of these basins come from the leftover planetesimals, which pushes them towards older ages. Bottke & Andrews-Hanna (2017) regard 4.1 Ga as both a tentative age of Nectaris basin and the timing of the origin of the late heavy bombardment (Marchi et al., 2012b; Morbidelli et al., 2012; Bottke et al., 2012, which is at odds with our chronology. For a more in-depth discussion regarding the pros and cons of the LHB model we refer to both Bottke & Norman (2017) and Mojzsis et al. (2019).

For Mercury we obtain $641^{+422}_{-442}$ basins from leftover planetesimals, a further $57^{+21}_{-22}$ from the E-belt, $4^{+4}_{-3}$ from the main belt and 125 from the comets. Conversely, the observed number of basins is about 40 (Werner, 2014). It is possible that most of the mercurian basins were also delivered while the innermost planet was in a magma ocean state, but from the inferred mercurian chronology shown in Figure 7 this would imply that its surface is no older than ca. 4.25 Ga. It is noteworthy that this is still somewhat older than the suggested resurfacing age of 4.1 Ga (Marchi et al., 2013) but generally much younger than the oldest surface ages of ca. 4.4 Ga for the Moon and Mars.

After the proposed late veneer impacts on the Earth (Brasser et al., 2016b) and Mars (Brasser & Mojzsis, 2017), which we argue occurred before ca. 4.43-4.48 Ga (Mojzsis et al., 2019), the expected largest object to strike the Moon has a diameter of $158^{+59}_{-43}$ km, on Mars it is $225^{+208}_{-105}$, $442^{+344}_{-212}$ on Earth and $352^{+306}_{-174}$ on Mercury. Most of these large impacts are expected to cause at least local crustal melting and regional-scale mechanical re-working (e.g. Abramov et al., 2013; Abramov & Mojzsis, 2016).

The diameter of the largest object to have struck the Moon both from leftover planetesimals and the E-belt is consistent with the suggested diameter of the South Pole-Aitken and Imbrium impactors. The decline of the leftover planetesimals is so rapid that >99.5% of the craters on the Moon resulting from the leftover population were created by 3.95 Ga, the suggested date of the formation of the Imbrium basin (e.g. Mojzsis & Ryder, 2001). The impact rate of E-belt objects on the Moon is much slower, however, which occurs with an e-folding time of 176 Myr. By 3.95 Ga, which is approximately 3 e-folding times after 4.5 Ga, some 95% of the craters on the Moon from the E-belt have been formed. Thus the Imbrium event occurring at 3.95 Ga cannot be ruled out as statistically anomalous at greater than 95% confidence. A similar argument applies to the lunar 4.25 Ga event reported in Apollo 14, 16 and 17 samples (Schäffer & Husain, 1974; Fischer-Gödde & Becker, 2012; Fernandes et al., 2013; Norman & Nemchin, 2014; Hopkins & Mojzsis, 2015; Kelly et al. 2018). We conclude that what has previously been interpreted as the late heavy bombardment (Ryder, 1990), was instead the consequence of events associated with the overall monotonic decline in the late accretion flux.

### 4.3 Atmospheric erosion from impacts

From the Monte Carlo simulations, we compute the fraction of the initial atmospheres of Venus, Earth and Mars that is lost due to impact erosion. We find that Earth lost about $34\%^{+22}_{-23}$, Venus lost approximately $50\%^{+33}_{-34}$ and Mars some $51\%^{+33}_{-35}$.

These estimates should be interpreted with great caution. First, we made crude assumptions about the primordial atmospheres of these bodies and different surface pressures or scale heights would yield different outcomes. Second, we do not take atmospheric replenishing from volcanism into account to determine the overall atmospheric loss.

That said, the loss on Earth and Venus are rather similar and thus we do not think that this mechanism can explain the differences in gas abundances between the atmospheres of Venus and Earth (Swindle, 2001). In the absense of superior atmospheric data during the earliest epochs of these planets we cannot conclude much more at this time, but we add that severe atmospheric erosion due to impacts remains a possibility.

## 5 Discussion

Since much of our methodology is based on the work of Brasser et al. (2016b) and Mojzsis et al. (2019) we refer to those studies for a discussion of the merits and shortfalls of our approach.

### 5.1 Chronology function

The martian chronology presented in Figure 6 has a poor agreement with the calibrated chronology presented by Werner et al. (2014); even though the slope (e-folding time) that we report from our dynamical simulations is similar to that of the calibrated Werner chronology, over most of the interval the computed value of $N_{20}$ is too low. Although our nominal density at 4.5 Ga is higher than that of Werner et al. (2014) by a factor of 1.5, the two curves only overlap within the $2\sigma$ uncertainties reported here for ages older than about 4.45 Ga; for younger ages the small deviation in the slopes causes ever greater discrepancy.

Despite the poor fit of our results to the Werner chronology, the dynamical chronology herein suggests that Mars' Hellas basin could be extremely ancient: Fassett & Head (2011) report that near Hellas basin $N_{20}=1.51\times10^{-4}$ km$^{-2}$, which, when solving for time using the Werner chronology, yields an age of 4.43 Ga; the corresponding age from the Ivanov chronology is 4.23 Ga and from our curve its age is also near 4.4 Ga. An independent age estimate can be derived from the work of Werner (2008), who reports $N_{10}=4.97\times10^{-4}$ km$^{-2}$ at Hellas basin. Assuming that for these crater diameters the crater size-frequency distribution scales as $D_{cr}^{-2}$ then the corresponding age of Hellas using the Ivanov chronology is nearly 4.2 Ga and as old as 4.39 Ga using the Werner chronology.

For the Moon the situation is similar: our dynamical and the Werner chronology are only within $2\sigma$ uncertainty for ages older than 4.45 Ga. At almost all times our lunar chronology curve is below the Werner curve. We have no obvious answer to explain this difference but will offer a few suggestions.

First, we consider the flux onto the moon relates to the flux onto the Earth by a factor of about 20 (simply the ratio of the impact probabilities) and have not derived other measures, although we corrected for the influence of the size-frequency distribution of the planetesimals in the Monte Carlo codes wherefrom we compute $N_{20}$. Second, in our dynamical models we did not take into account the collisional evolution of the planetesimals. Omitting this from our models is a fair point of critique. Collisional evolution substantially increases the initial rate of elimination of most planetesimals and thus increases the slope of the curves (Dohnanyi, 1969; Durda & Dermott, 1997; Bottke et al., 2007). The study by Bottke et al. (2007) concluded that the probability of producing Nectarian and/or early-Imbrian era basins from leftover planetesimals when accounting for collisional evolution drops below the 3σ level. As such, collisional evolution is working against what we try to achieve here. The interpretation of collisional evolution nonetheless requires appropriate prudence. From the rotation period distribution of large asteroids Pravec et al. (2002) suggested that asteroids larger than ~40 km in diameter have a distribution close to Maxwellian, which would mean that they are either original bodies of the asteroid main belt or its largest, collisionally evolved remnants.

Rather than collisions speeding up the removal rate of material, what is instead required is a mechanism to *decrease* the average impact rate for the first billion years from leftover planetesimals. Since the dynamical evolution of small and large planetesimals is essentially the same, such a slowing of the impact rate can only be achieved by varying the architecture of the inner planets. This is a tall, neigh impossible order. We are not of the opinion that *ad hoc* collisional evolution will resolve the discrepancy between our suggested lunar chronology and those of Neukum and Werner.

A third consideration pertains to the contribution from the E-belt. It is not clear whether or not this reservoir actually existed, and if it did it was either a natural extension of the asteroid belt or part of the reservoir of leftover planetesimals. Inspection of the Grand Tack simulations of Brasser et al. (2016a) indicate that at their end there are a few planetesimals which occupy the E-belt region, so that the latter could be just part of the leftover planetesimals; the mass implanted in that region during in the Grand Tack model (0.16% of the initial population or roughly $5\times10^{-3}$ $M_E$) turns out to be higher than the initial mass inferred by Bottke et al. (2012) by a factor of ~10. A different dynamical model, that starts with the asteroid belt completely devoid of material (Raymond & Izidoro, 2017), favours the hypothesis that the E-belt region is the result of implantation by planetesimals from the Earth-Venus region; further study is needed to determine whether the E-belt was a natural extension of the asteroid belt or is in fact implanted. The Hungaria group of asteroids may provide insight as spectroscopically they resemble the aubrite class of meteorites (e.g. Kelley & Gaffey, 2010) and thus may favour an inner solar system origin. In any case, subtracting the E-belt contribution will make the lunar and martian chronologies proposed here results in a product that is even less consistent with the calibrated chronologies of Neukum and Werner, because it is the only reservoir whose dynamical half-life is comparable to the Werner (and Neukum) chronologies.

If we were to impose that most late accretion comes from leftover planetesimals (Morbidelli et al., 2018), then we cannot fit the lunar impact curve through most of the Neukum calibration points, not even when we restrict ourselves to data points younger than 4.1 Ga or even 3.9 Ga. Morbidelli et al. (2018) consider that chronology calibration constraints older than 3.9 Ga are unreliable and these were thus ignored. As such, they were satisfied with a chronology curve that was steeper than Neukum's for ages older than 4 Ga. In our study, we give weight to older data points, up to 4.35 Ga. These additional data points change the shape of the chronology function by making it shallower.

Any attempt to restrict the chronology curve to leftover planetesimals requires a leftover mass that is 300 times as much as we calculated in section 4.1. This would not only violate constraints from the Hadean zircons, but it is also hopelessly inconsistent with the outcome of dynamical simulations of terrestrial planet formation. If the mechanism of Zhu et al. (2019) is invoked then such a high mass increase for the leftover population requires that the lunar HSEs are only retained after 4.1 Ga.

The chronology function for the leftover planetesimals is much steeper than what was shown in Brasser et al. (2016b). The reasons are twofold. First, Brasser et al. (2016b) fitted the decline of the total population rather than the impacts onto the planets. Second, their simulations were run between 4.4 Ga and 4.25 Ga, while the ones presented here were run from 4.5 Ga to 3.5 Ga. Therefore, the range over which the chronology is fitted in this work is much longer. The decline in the leftover planetesimals is steep, but the e-folding time in the second term is consistent with the delivery of Earth crossers (Migliorini et al., 1998), while the e-folding time of the third term is consistent with the injection and elimination of Mars crossers (Michel et al., 2000), lending some credibility to the fit.

Taken together, our simulations indicate that the measured crater chronologies seem to require the presence of the E-belt reservoir.

**5.2 The mass of the E-belt**

If we accept the argument that the E-belt reservoir may hold the key to the crater chronology of the inner planets, then we can compute the desired initial mass increase for its impact profile to be consistent with the lunar and martian Werner chronologies. At 4.5 Ga the lunar Werner chronology is computed to be $N_{20}=10^{-3}$ km$^{-2}$, while our calculated E-belt contribution gives $0.237\times10^{-3}$ km$^{-2}$ (Table 4), requiring a mass increase of 400% to be fully compatible with the Werner chronology. For Mars Werner et al. (2014) compute $N_{20}=0.489\times10^{-3}$ km$^{-2}$ at 4.5 Ga and our E-belt contribution is $0.160\times10^{-3}$ km$^{-2}$, necessitating a mass increase of 300%. In the case of the Moon, such a factor of 4 mass increase would miss the calibration points from Neukum (1983) and the ones we created in this work. To proceed, we do a simple experiment wherein the flux is dominated by leftover planetesimals and the E-belt. The total chronology-defining flux was assumed to be of the form $N_{20}(t)=A\times f_L(t)+B\times f_E(t)$, where $f_L(t)$ and $f_E(t)$ are the cratering chronology functions provided in equation (4) for the leftovers and the E-belt, and A and B are free parameters. A best fit through the calibration points of Neukum (1983) results in A~0 and B~10, while a fit through the calibration points introduced in this work results in A~0 and B~4 (irrespective of whether the Nectaris point is included). Performing the same experiment for Mars we once again obtain A~0 and B~4. Forcing A=1 does not change the value of B in a substantial way. These combined fits suggest roughly a factor of four increase in the mass of the E-belt, although it can be as high as ten. We have plotted this in Figure 8.

A possible argument in favour of such a mass increase for the E-belt comes from Vesta. In section 4.1 we argued that the amount of mass from the E-belt striking Vesta could produce the massive Rheasilvia basin near Vesta's south pole if all of it came in just a single impact with a planetesimal whose diameter is 66 km. At present there are 500 bodies with D>62 km in the asteroid belt, and 360 with D>77 km (Jedicke et al., 2002; Bottke et al., 2005). The current main belt mass is comparable to the initial mass estimated for the E-belt (Bottke et al., 2012). The probability that one of these large objects strikes Vesta is approximately $10^{-4}$, so with the current asteroid belt population the chance of creating such a basin on Vesta is about 5%, or a 2$\sigma$ event. If we raise the initial mass of the E-belt then the creation of Rheasilvia over the lifetime of Vesta increases to 20%, or even higher when the decline in the mass of the main belt is considered.

A further test comes from the ages of the youngest lunar basins. We can use the chronology of the E-belt to calculate the expected age of the youngest lunar basin. We create on average $16^{+8}_{-8}$ lunar basins from the E-belt, so this would increase to $64^{+32}_{-32}$ if we quadruple the mass of the E-belt. The E-belt lunar chronology is a simple exponential $f_E(t)=e^{-t/\tau}$ and thus setting $f_E(t)=1/64$ and solving for $t$ we obtain get $t=729^{+71}_{-121}$ Myr, so that the youngest lunar basin is expected to be $3771^{+121}_{-71}$ Ma. Within uncertainty this corresponds to the age of Orientale basin.

It turns out that our suggested higher initial mass for the E-belt poses a potential problem for the current Hungaria asteroid population (Warner et al., 2009). Bottke et al. (2012) conclude that the Hungaria population currently contains 4±2 objects large enough to form a basin on the Moon. Bottke et al. (2012) finds a typical depletion factor of 200-1000 for the E-belt over 4.1 Gyr. We have only run our simulations for a model time of 1 Gyr so we use the population decline rate of Bottke et al. (2012) to compare our results to that of the Hungaria population. From their Figure 2 we estimate that extending the E-belt simulations by another 400 Myr would result in a total *expected* depletion after 4.5 Gyr of evolution that is a factor of two greater than that suggested by Bottke et al. (2012). To match the current Hungaria population thus requires a doubling of the initial mass of the E-belt at 4.5 Ga from that of Bottke et al. (2012). Here, we advocate a factor of four increase, which is within the range of uncertainties, and is purely caused by giant planet migration commencing earlier than in Bottke et al.

(2012). That said, the uncertainty at 4.1 Ga in the decay curves presented in Bottke et al. (2012) are large enough that a factor of a few increase in the mass of the E-belt is still consistent with the current Hungaria population.

Raising the E-belt's mass by a factor of four to five to match its impact contribution to the Moon and Mars with the published crater chronologies appears consistent with dynamical simulations of terrestrial planet formation and possibly with the current Hungaria population, but it poses a different problem. Such a mass increase would raise the value of $N_{20}$ from the E-belt alone to $\sim 10^{-3}$ km$^{-2}$. From our Monte Carlo simulations this crater density corresponds to a total mass augmentation to the Moon of approximately 0.025 wt.%, which is the value deduced from its HSE abundances (Day et al., 2016), and would imply that there is little to no room for any contribution from leftover planetesimals. The lunar tungsten isotopes, on the other hand, suggest a late addition of 0.05 wt.% (Touboul et al., 2015; cf. Thiemens et al., 2019) so that about half of it could come from the leftover planetesimals. These results raise the possibility for an even more diminished role of leftover planetesimals. Simulations of terrestrial planet formation – with collisional evolution taken into account – hint at the possibility of a very few small planetesimals remaining in the system (Walsh & Levison, 2016). Yet, it is not clear to us whether the total mass in planetesimals – which is dominated by the large members – is also much reduced.

A further complication may stem from the current population of large asteroids in the main belt. Nesvorný et al. (2017a) obtain a similar rate of decline of the E-belt and impact probability as we do. They use the current number of large asteroids in the main belt as their normalisation for the initial population. From this they conclude that the total number of lunar impactors with $D_i>10$ km from the E-belt and main asteroid belt should have been between 4 and 10, which translates into 400 to 1000 impacts for planetesimals with $D_i>1$ km; that is a factor of 10-20 lower than what we calculated in this work. The reason for this discrepancy is not clear to us, but we speculate that it is the either result of the different normalisations (large asteroids versus lunar craters and HSE abundances) or due to the differences in the employed size-frequency distributions. In any case, their work is possibly inconsistent with the Werner and Neukum chronologies, while ours is possibly inconsistent with the number of large asteroids in the main belt.

Both further measurements of the lunar HSE abundance and its inferred amount of late accretion plus improved dynamical models of terrestrial planet formation and subsequent evolution of the leftover planetesimals and the asteroid belt are needed to resolve the dilemmas mentioned above.

**5.3 The age of the oldest surfaces**

Given the chronologies shown in Section 4 a question arises: how old are the oldest surfaces on the Moon and Mars that we can date from crater counts? Crater saturation occurs when the density of craters is $N_{eq}=0.91D_{cr}^{-2}$ km$^{-2}$ (Gault, 1970) and thus for 1 km craters $N_{eq}=0.91$ km$^{-2}$. In reality, however, saturation is already reached when $logN_S = -1.33 - 1.83 logD_{cr}$ (Hartmann, 1984) which is about $N_1 \sim 0.05$ km$^{-2}$ and $N_{20}=1.95\times10^{-4}$ km$^{-2}$; actually, the observed saturation occurs at $(0.01-0.1)N_{eq}$ (Richardson, 2009), and it has been suggested that the lunar highlands, with a maximum $N_{20}=2.8\times10^{-4}$ km$^{-2}$, are in such saturation (Head et al., 2010; cf. Xiao & Werner, 2015). Taking the saturation of $N_1$ to be 0.05 km$^{-2}$, the lunar and martian surfaces reach theoretical saturation around 4.1 Ga and 4.2 Ga. This is much younger than documented lunar and martian zircon ages of ca. 4.42 Ga; these zircons likely indicate crust formation, or at least crust preservation. Using the lunar basins as the chronometer, however, alters the story somewhat and predicts an older lunar surface age.

The saturation limit for basins using the formula from Hartmann (1984) is $N_{eq}=1.37\times10^{-6}$ km$^{-2}$, which is slightly higher than the observed value of $N_{300}=1.1\times10^{-6}$ km$^{-2}$ reported by Neumann et al. (2015) from the GRAIL mission data. For every basin there should be an additional ~556 craters with diameter 20 km or greater if the projectiles had the same size-frequency distribution as the main asteroid belt. The saturation density for these smaller craters when extrapolating from that of the basins is $N_{eq}=7.6\times10^{-4}$ km$^{-2}$, which, when reading from Figure 5 and multiplying by 400, corresponds to a surface age of about 4.45 Ga using Figure 7. Repeating this calculation using the alternative saturation limit $N_s=0.05N_{eq}=5.1\times10^{-7}$ km$^{-2}$ for the basins yields a lunar surface age of 4.3 Ga. Thus, we run into the problem pointed out by Werner (2014) that the basin record predicts a surface age that is about 150-200 Myr older than that computed from the smaller craters; the maximal lunar surface age calculated here, however, is consistent with that of Werner (2014), and the lunar bombardment has been in steady, monotonic decline ever since. A similar argument applies to Mars, and possibly to Mercury, with portions of the martian surface age calculated from the basins potentially reaching 4.45 Ga. The Frey (2008) basin density on Mars, however, appears to exceed the saturation limit. We argue that the number of buried martian basins could exceed 30 but cannot be much higher than ca. 100.

Observations of craters on the Moon, Mars and Mercury suggest that, although still debated, the comets play an inferior role to inner Solar System projectiles in the bombardment of the inner Solar System (Neukum & Ivanov 1994, Werner et al., 2002, Strom et al., 2005, Rickman et al., 2017, Morbidelli et al., 2018). This conclusion is supported by Figures 5 to 7 if the fading law that we have adopted from Nesvorný et al. (2017b) is valid.

We now address the meaning of model ages for the Moon. In Figure 5 we added the calibration observations from Neukum (1983). An important point to emphasize here is that the idea behind the calibration is that the rocks are dated with U-Pb or $^{40-39}$Ar chronology, which provide an apparent age. Performing crater counts near where the rock was collected should then link the rock age to the crater density. In practise, however, the Moon undergoes a nevertheless rather complex geological evolution including impact gardening, wherein rocks bounce around the surface from repeated impacts over lunar history. For example, lunar impact breccia 14311 is characterised by impact and deposition in the Imbrium ejecta blanket at ca. 3.9 Ga, followed by a second impact event at ca. 110 Ma, and limited solar heating to less than 2.0 Myr (Kelly et al., 2018). If this sequence of events for sample 14311 is not atypical for other lunar samples then it is unclear whether the $^{40-39}$Ar ages, and potentially the U-Pb ages too, that were measured from these samples correspond to the local geology where they were found; long distance associations are even less straightforward. For example, when discussing lunar breccia 67955 Norman et al. (2014) conclude that the petrological and geochemical characteristics of the 67955 noritic anorthosite component suggest that it formed by an impact in the Procellarum-KREEP Terrane, and was transported to the Apollo 16 site interpreted to be as Imbrium ejecta rather than ejecta from the nearby Nectaris basin. Documented $^{40-39}$Ar ages of 67955 and related Apollo 16 samples all cluster near 3.95 Ga rather than the proposed age of 4.1-4.2 Ga for Nectaris basin (Fischer-Gödde & Becker, 2012; Norman & Nemchin, 2014). The most parsimonious approach to obtain (improved) absolute ages is to tie such ages to specific local events. On Mars, Werner et al. (2014) tied the origin of the shergottite-type and ALH84001 meteorites to the Mojave impact crater which allowed them to obtain an absolute martian chronology (see also Werner, 2019). We have crudely attempted to do the same on the Moon in this work for the three calibration points reported above. The preponderance of reset ages near 3.9 Ga has been commonly attributed to contamination of samples with Imbrium ejecta based on its chemical affinity (Norman et al., 2010), and a similar high number of ages near 4.2 Ga should also be considered to coincide with a large impact event on the Moon, possibly the formation of Nectaris (e.g. Hopkins & Mojzsis, 2015); however, for these older ages no firm candidate impact basin has yet been identified.

## 6 Conclusions

In this work we have computed mass flux functions for late accretion to the terrestrial planets and the Moon, and made estimates of the total amount of mass accreted by dwarf planets Vesta and Ceres from three sources: planetesimals left over from terrestrial planet formation, the hypothesised E-belt, and comets. We attempt to create reliable and ultimately, testable estimates of the amount of late accretion from comets to the terrestrial planets. We further demonstrate that Mars and the Moon accrete a proportionally large amount of cometary material compared to Venus and the Earth. A hint of this cometary accretion may be present in lunar D/H values and Xe isotopes (Greenwood et al., 2011; Bekaert et al., 2017), and in terrestrial Xe (Marty et al., 2017).

We show that the background mass augmentation from small planetesimals to the Earth and Mars is far lower than the estimates derived from HSE abundances in their mantles. This supports the idea that both planets were struck by anomalously large singular objects early on, but nevertheless long after their formation, rather than by a swarm of smaller objects.

We use the impacts recorded onto the planets from dynamical simulations to calculate the impact chronology curves for the Moon, Mars and Mercury rather than relying on the decline of the population as a whole. The dynamical lunar chronology is different from the chronologies of Werner and Neukum, while for Mars it is a reasonably good match to the calibrated Werner chronology; for Mercury no known samples exist and all curves are purely theoretical. Quadrupling the mass of the E-belt reservoir makes the dynamical lunar and martian chronologies be consistent with the calibrated Werner chronologies; an order of magnitude higher mass is needed to match the calibration points of Neukum (1983), but this violates constraints from the Hungaria population, lunar HSE abundances and possibly its tungsten isotopes. We further show that with our current knowledge of solar system evolution the dynamically derived lunar and martian chronologies do not quite match each other, and strongly caution against simple extrapolation of the crater chronology from one body to the next.

We end with one prediction from this work. If our dynamical chronology model is correct then the lunar HSE abundances and Os isotopes should reflect the composition of the material that made up the leftover planetesimals and the E-belt. Fischer-Gödde & Becker (2012) suggest that late accretion to the Moon consisted of a mixture of material akin to enstatite and carbonaceous chondrites. If future measurements confirm these results, then it may be possible that a 'carbonaceous chondrite' signal may instead be a remnant of the very early cometary contribution.

**Acknowledgements**

The authors thank Vera Assis Fernandes for valuable discussions. Constructive criticism from editor Alessandro Morbidelli, as well as Bill Bottke and an anonymous reviewer substantially improved the quality of the manuscript. RB acknowledges financial assistance from the Japan Society for the Promotion of Science (JSPS) International Joint Research Fund (JP17KK0089) and JSPS Shingakujutsu Kobo (JP19H05071). All authors thank the Collaborative for Research in Origins (CRiO) at the University of Colorado, which is supported by The John Templeton Foundation: the opinions expressed in this publication are those of the authors, and do not necessarily reflect the views of the John Templeton Foundation. SCW is grateful for financial assistance from the Research Council of Norway through the Centre of Excellence funding scheme, project number 223272 (Centre for Earth Evolution and Dynamics). The authors also thank the Ph.D. students of the graduate planetary course ASTR/GEOL 5835 "Late Accretion" at the University of Colorado, Boulder who participated in discussions with us during the formulation of this study. The source code for the Monte Carlo simulations is available upon request from RB.

**Appendix**

| Planet | $\langle p_{imp}\rangle$ [%] | $\langle v_{imp}\rangle$ [km s$^{-1}$] | Expected mass striking the surface [ppm] |
|---|---|---|---|
| Mercury | $5.2^{+1.4}_{-1.5}$ | $36^{+24}_{-20}$ | $919^{+1280}_{-734}$ |
| Venus | $17^{+3.7}_{-3.1}$ | $26^{+15}_{-12}$ | $170^{+177}_{-130}$ |
| Earth | $13^{+2.5}_{-2.0}$ | $22^{+13}_{-9}$ | $115^{+130}_{-89}$ |
| Moon | $0.59^{+0.12}_{-0.20}$ | $19^{+15}_{-12}$ | 250 (fixed) |
| Mars | $0.85^{+0.13}_{-0.17}$ | $15^{+10}_{-9}$ | $79^{+139}_{-69}$ |
| Vesta | $5.4\times10^{-4}$ | 13 | 534 |
| Ceres | $7.6\times10^{-4}$ | 13 | 710 |

Table A1: Impact probability, average impact velocity and typical mass augmentation for all inner Solar System planets from planetesimals left over from primary accretion.

| Planet | $\langle p_{imp}\rangle$ [%] | $\langle v_{imp}\rangle$ [km s$^{-1}$] | Expected mass striking the surface [ppm] |
|---|---|---|---|
| Mercury | $0.90^{+0.15}_{-0.03}$ | $41^{+27}_{-24}$ | $71^{+128}_{-56}$ |
| Venus | $5.4^{+0.7}_{-1.0}$ | $26^{+19}_{-14}$ | $28^{+39}_{-17}$ |
| Earth | $5.5^{+0.7}_{-1.0}$ | $21^{+14}_{-9}$ | $24^{+35}_{-15}$ |
| Moon | $0.25^{+0.03}_{-0.04}$ | $18^{+16}_{-12}$ | $93^{+154}_{-80}$ |
| Mars | $2.2^{+0.1}_{-0.2}$ | $11^{+11}_{-5}$ | $94^{+155}_{-67}$ |
| Vesta | 0.016 | 9 | 1500 |
| Ceres | 0.0022 | 11 | 56 |

Table A2: Impact probability, average impact velocity and typical mass augmentation for all inner Solar System planets from planetesimals from the E-belt with its canonical mass obtained in Bottke et al. (2012).

| Planet | $\langle p_{imp}\rangle$ [$10^{-4}$] | $\langle v_{imp}\rangle$ [km s$^{-1}$] | Expected mass striking the surface [ppm] |
|---|---|---|---|
| Mercury | $4.7^{+1.5}_{-3.3}$ | $43^{+27}_{-32}$ | $2.1^{+3.4}_{-2.0}$ |
| Venus | $15^{+2.3}_{-1.5}$ | $31^{+16}_{-17}$ | $0.74^{+1.2}_{-0.67}$ |
| Earth | $17^{+4.2}_{-3.6}$ | $25^{+10}_{-12}$ | $0.59^{+1.1}_{-0.53}$ |
| Moon | $0.75^{+0.65}_{-0.43}$ | $22^{+11}_{-15}$ | $1.8^{+2.8}_{-1.7}$ |
| Mars | $7.1^{+3.2}_{-2.2}$ | $13^{+9}_{-7}$ | $2.1^{+3.8}_{-2.0}$ |
| Vesta | 1.1 | 5.3 | 920 |
| Ceres | 3.4 | 5.6 | 774 |

Table A3: Impact probability, average impact velocity and typical mass augmentation for all inner Solar System planets from planetesimals from the main asteroid belt (A-belt).

| Planet | $\langle p_{imp} \rangle$ [x $10^{-6}$] | $\langle v_{imp} \rangle$ [km s$^{-1}$] | Expected mass striking the surface [ppm] |
|---|---|---|---|
| Mercury | $0.13^{+0.22}_{-0.08}$ | $34^{+10}_{-5}$ | $42^{+14}_{-13}$ |
| Venus | $1.9^{+2.9}_{-1.3}$ | $26^{+3.8}_{-2.5}$ | $42^{+7}_{-7}$ |
| Earth | $3.1^{+3.3}_{-2.8}$ | $22^{+2.9}_{-1.7}$ | $57^{+8}_{-8}$ |
| Moon | $0.16^{+0.20}_{-0.08}$ | $20^{+3.7}_{-1.9}$ | $239^{+64}_{-70}$ |
| Mars | $2.5^{+3.3}_{-1.2}$ | $13^{+1.9}_{-0.6}$ | $409^{+59}_{-42}$ |
| Vesta | 0.0093 | 11 | 3900 |
| Ceres | 0.025 | 11 | 2900 |

Table A4: Impact probability, average impact velocity and typical mass augmentation for all inner Solar System planets from comets.

**Figures**

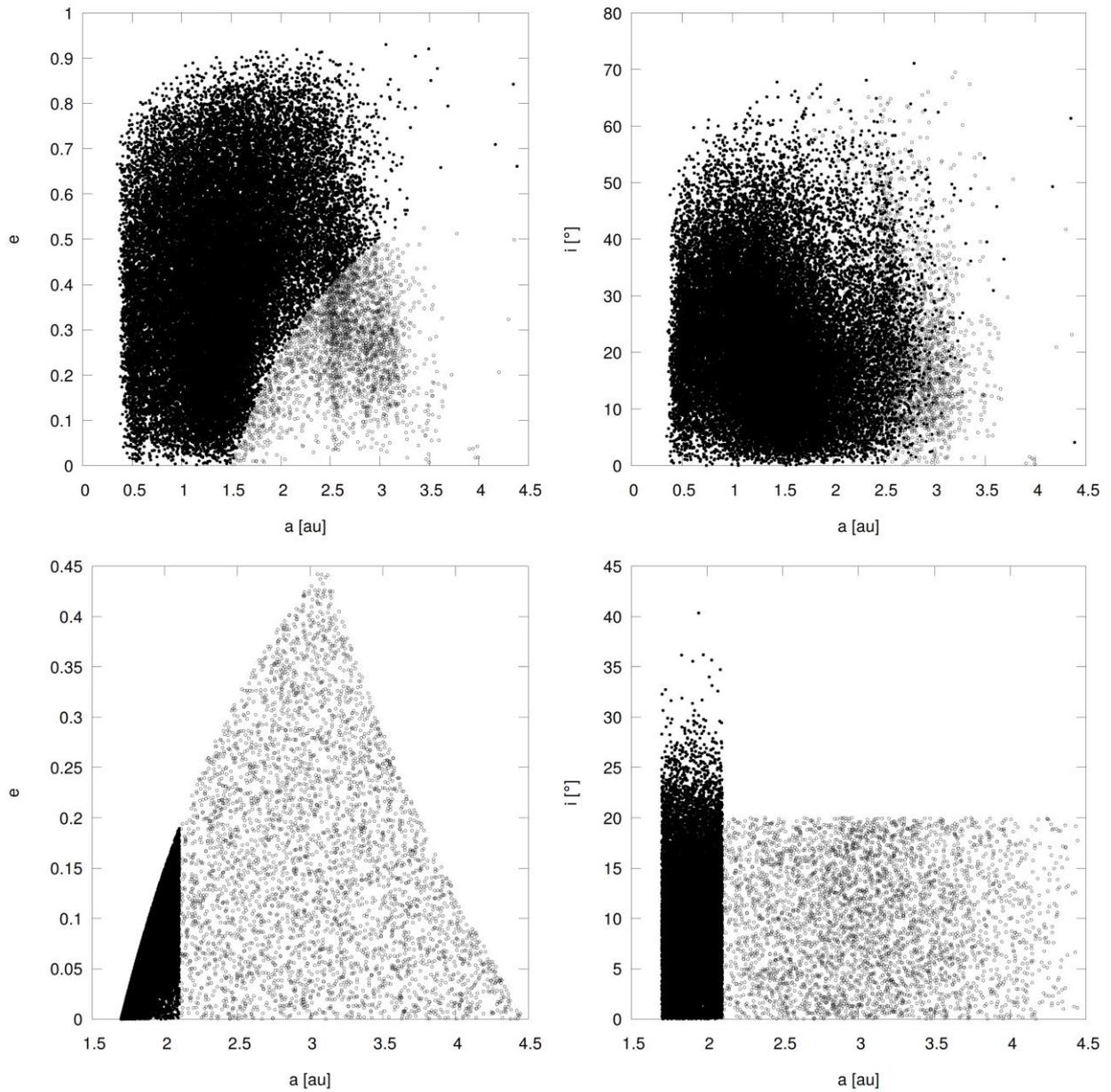

Figure 1: Initial conditions for the leftover planetesimals (top panels) and the E-belt and A-belt (bottom panels). Right column: eccentricity versus semi-major axis. Right column: inclination versus semi-major axis. In the top panels the open symbols are leftover planetesimals that were not included in the simulations. In the bottom panels the open symbols denote the A-belt while the filled symbols denote the E-belt.

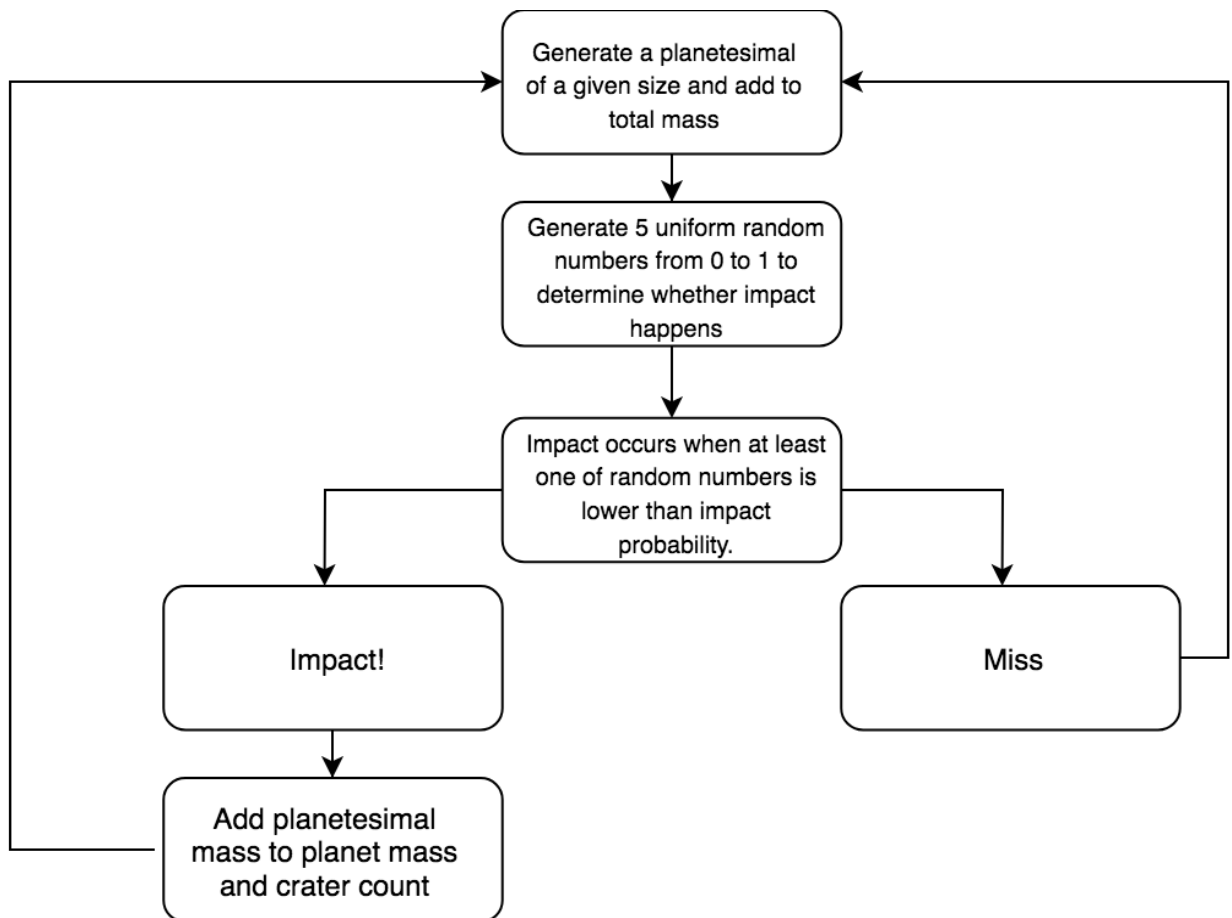

Figure 2: Flowchart of the Monte Carlo simulations.

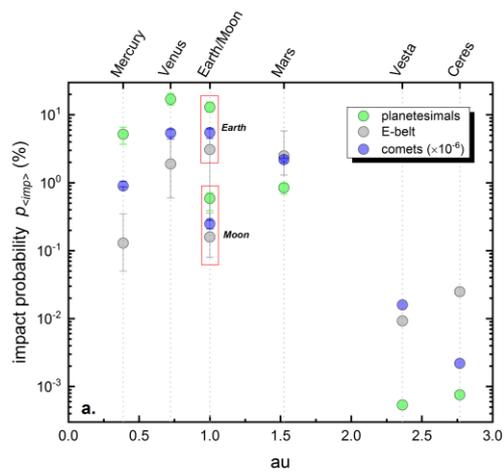

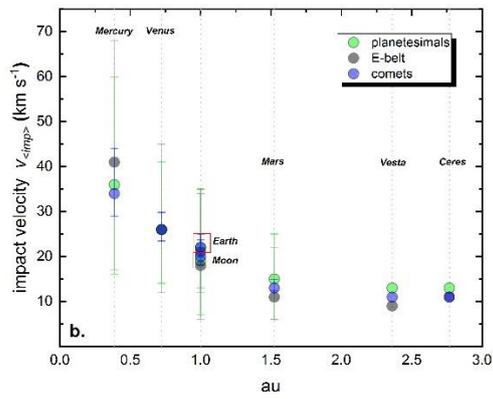

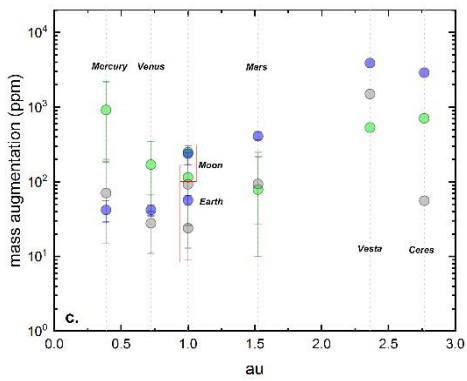

Figure 3. Impact probabilities (a), impact velocities (b) and mass augmentations (c) for the inner planets, Vesta and Ceres for the three sources of impactors: leftover planetesimals, E-belt and comets.

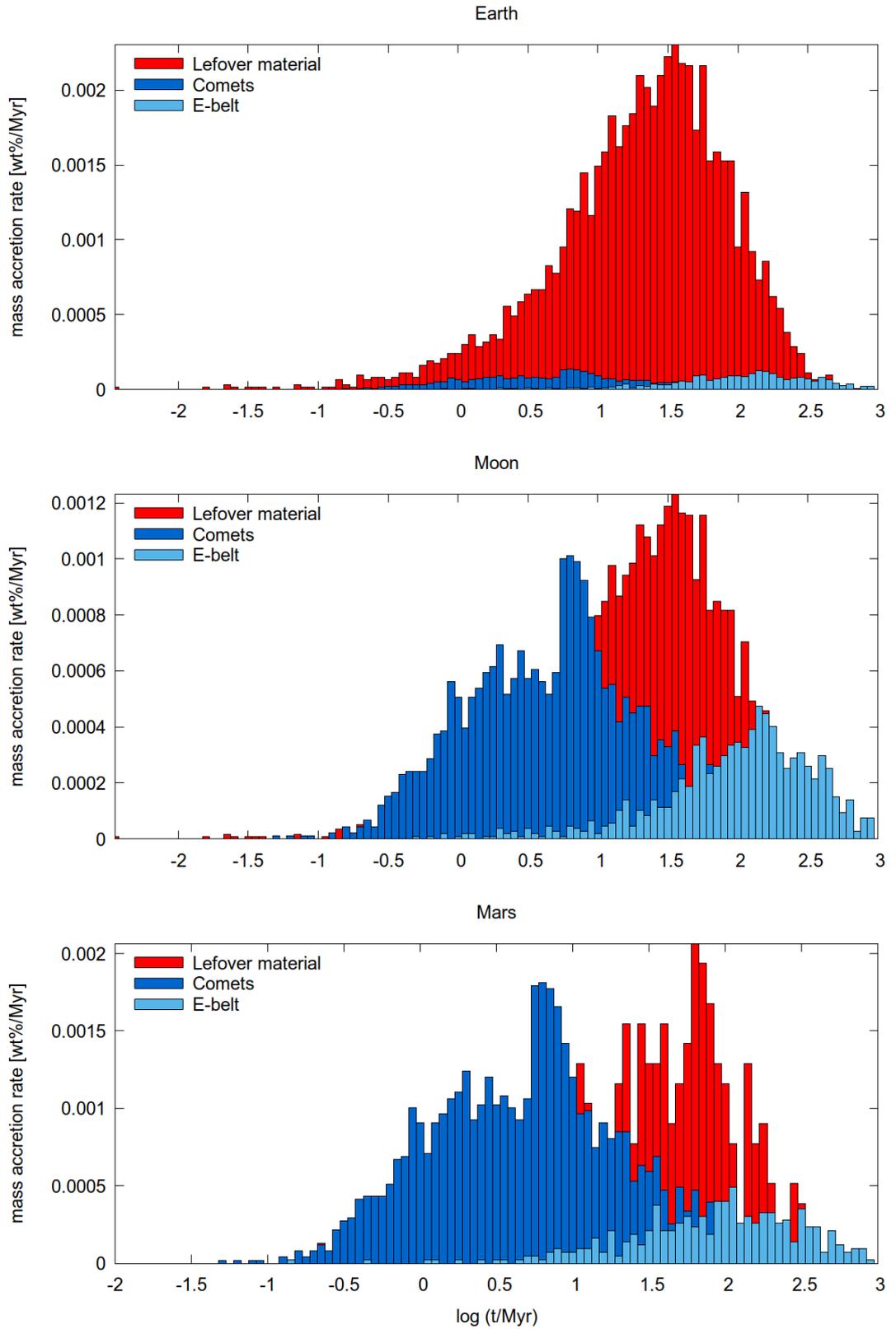

Figure 4: Rate of accretion onto the Earth, Moon and Mars from the three major sources (comets, leftover planetesimals and E-belt).

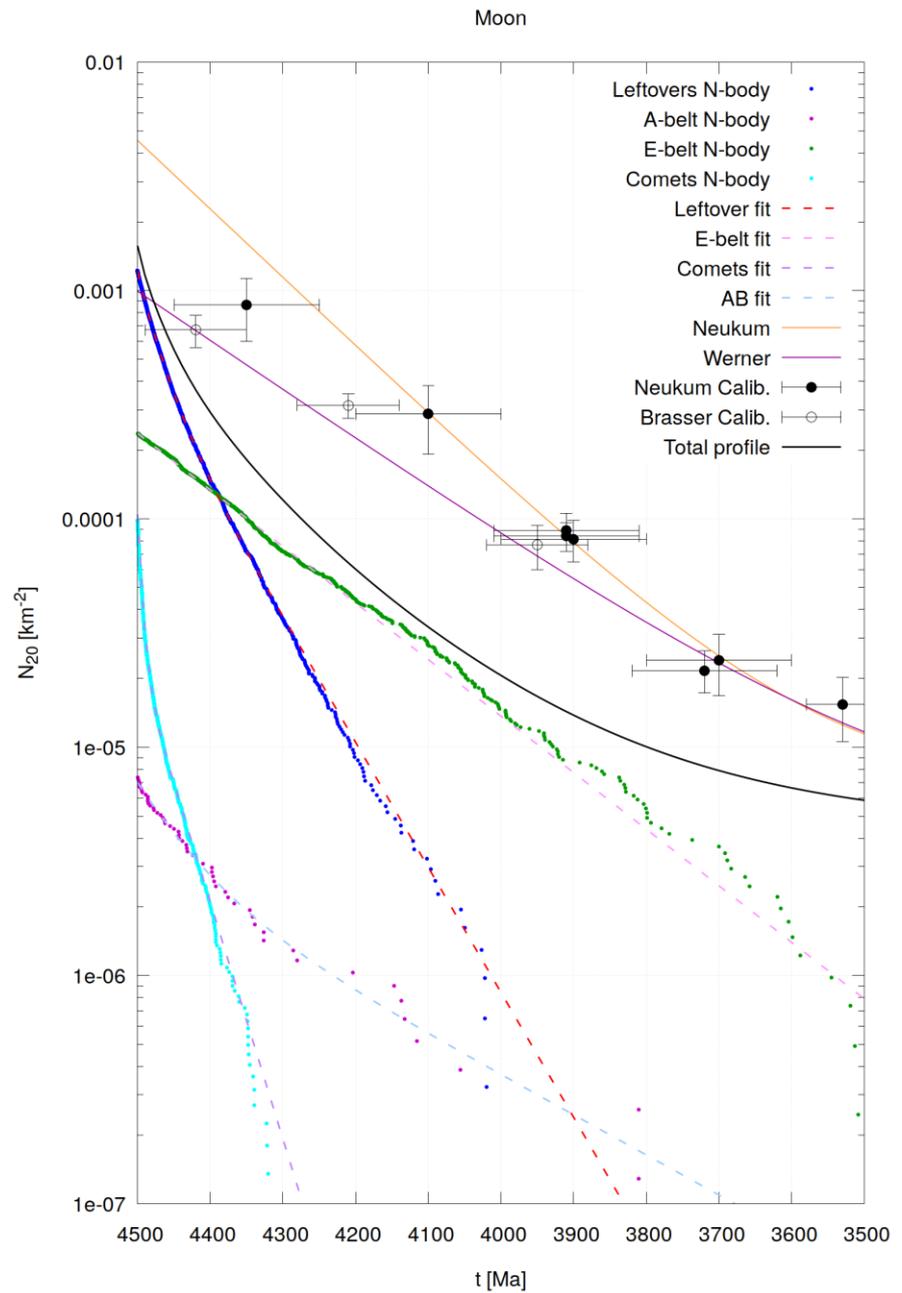

Figure 5: Lunar chronology from a combination of dynamical and Monte Carlo simulations. The three reservoirs are the comets, leftover planetesimals and the E-belt with the best fits. The calibrated Werner and Neukum chronologies are shown separately as solid lines. The filled circle data points are calibrations from Neukum (1983). The open circles are an attempt by us at a new, albeit somewhat provisional, calibration.

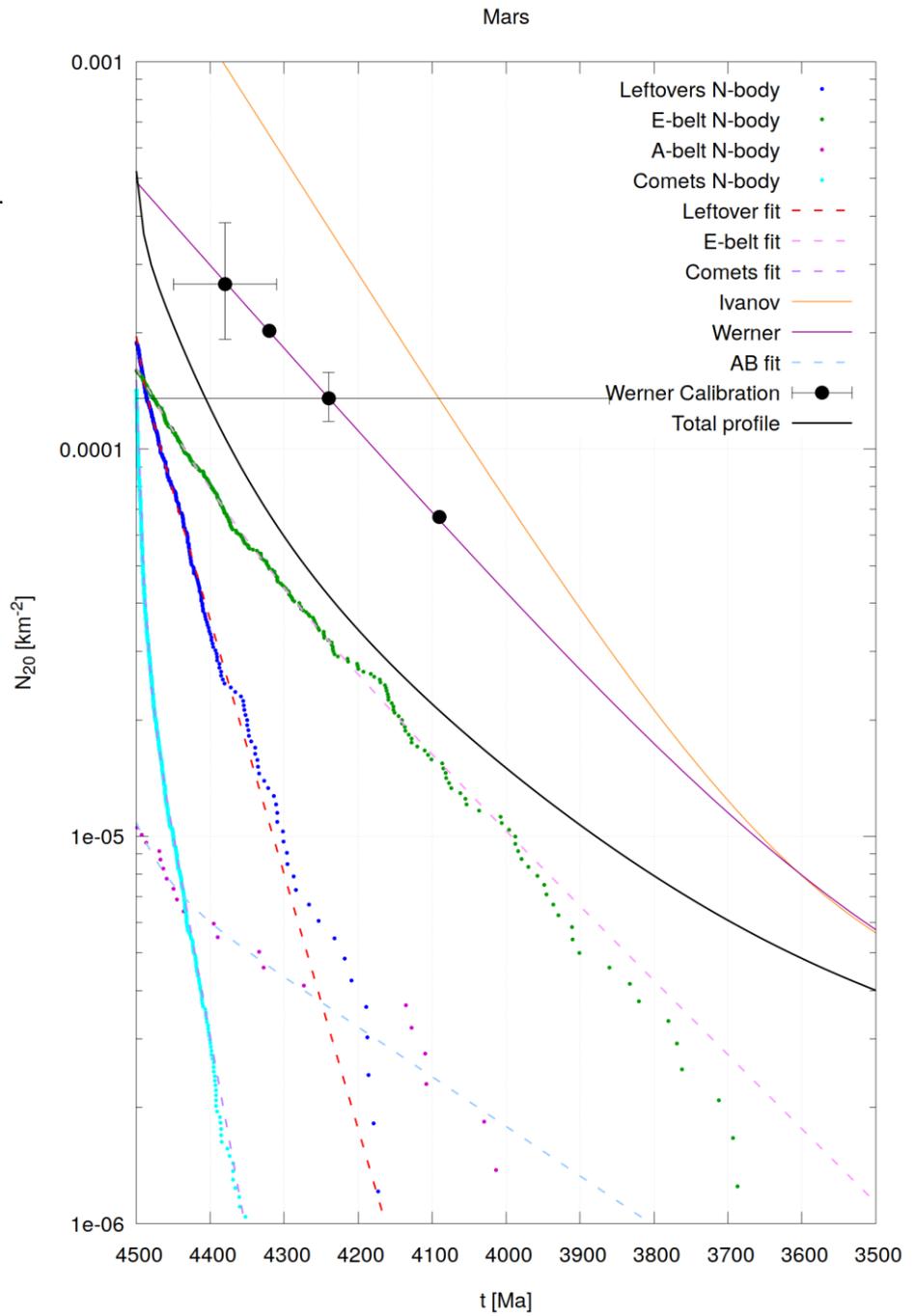

Figure 6: Martian chronology from a combination of dynamical and Monte Carlo simulations. The three reservoirs are the comets, leftover planetesimals and the E-belt with the best fits. The Werner and Ivanov chronologies are shown separately as solid lines.

Figure 7: Mercurian chronology from a combination of dynamical and Monte Carlo simulations. The three reservoirs are the comets, leftover planetesimals and the E-belt with the best fits. The Werner and Neukum chronologies are shown separately as solid lines.

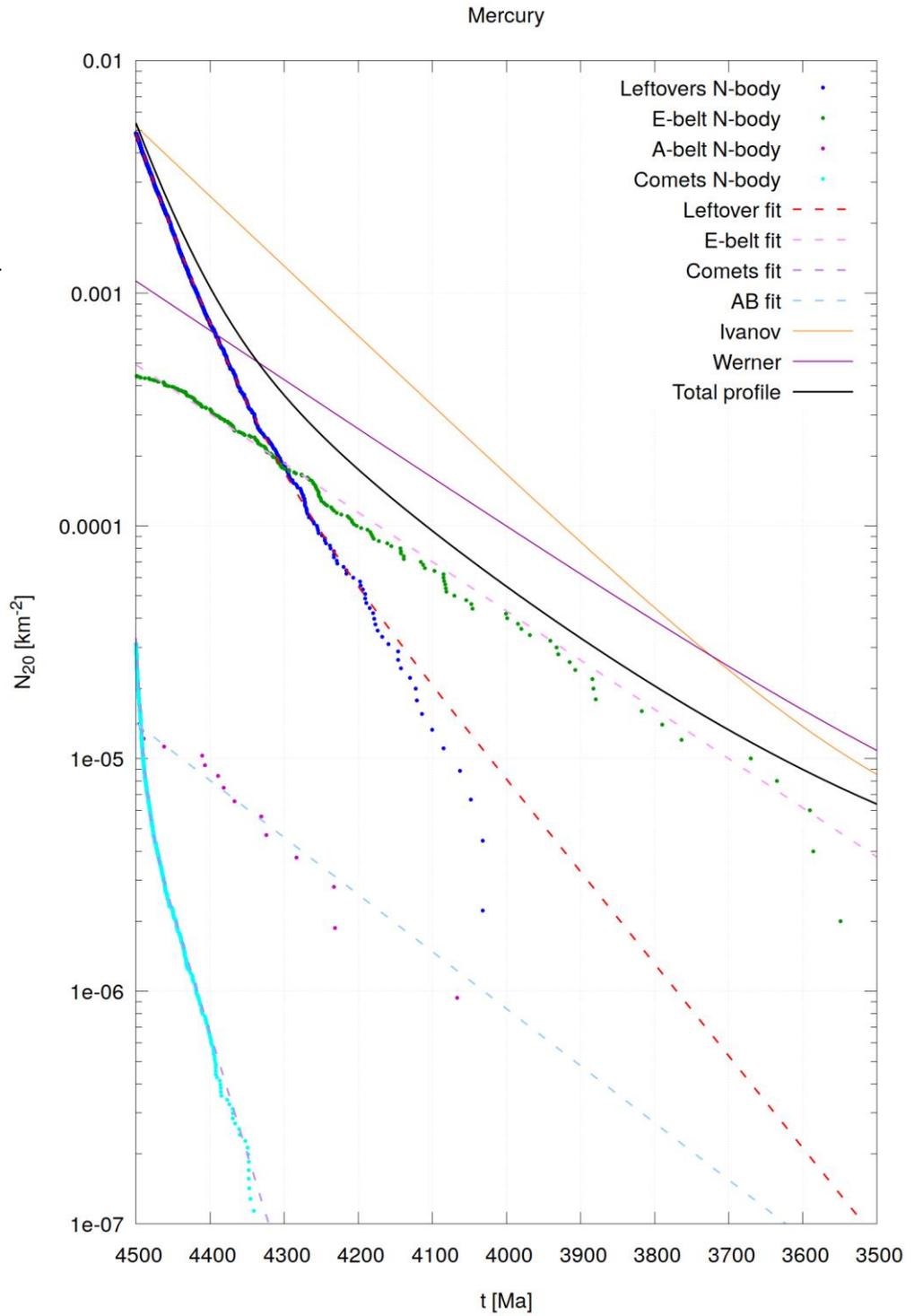

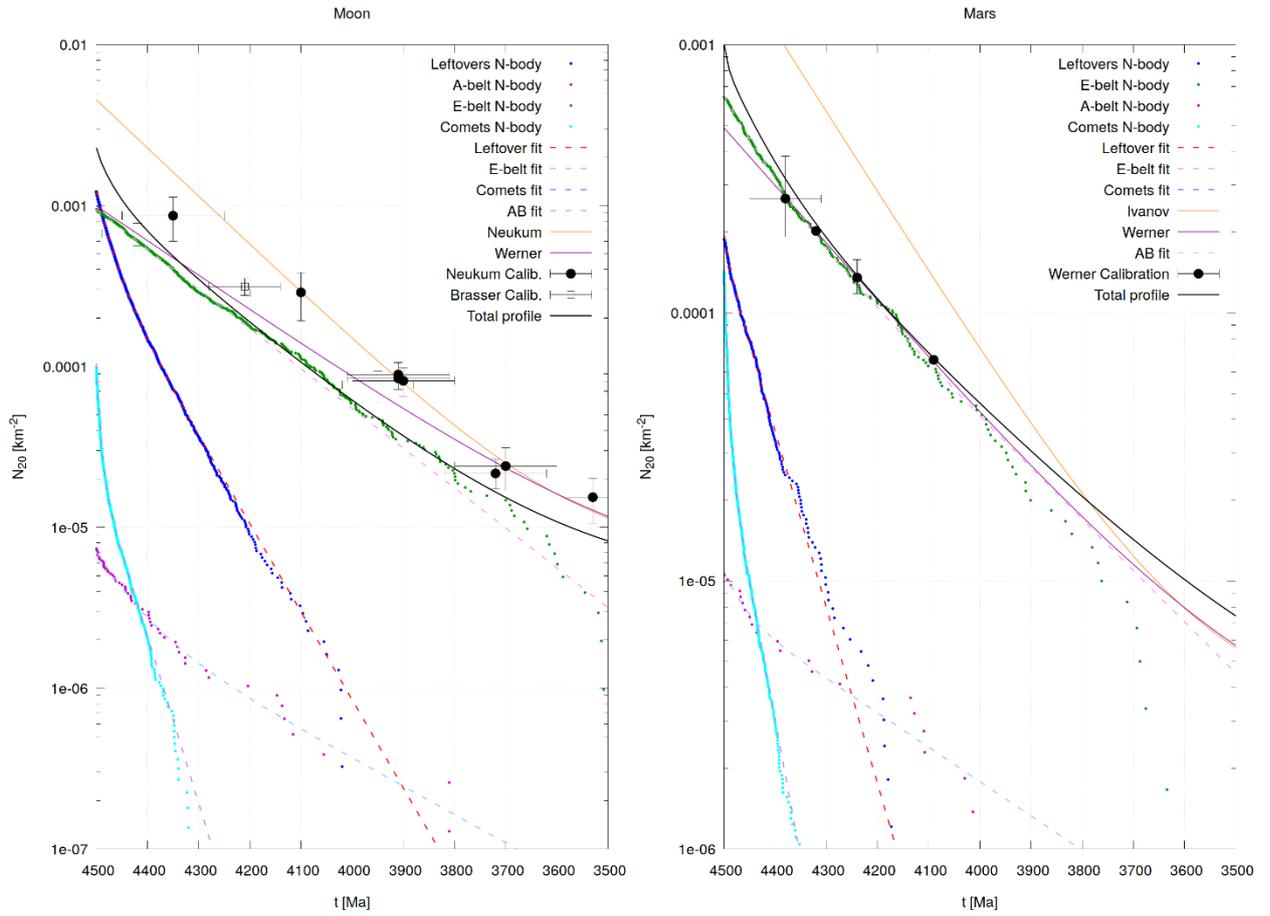

Figure 8: Lunar (left) and martian (right) dynamical chronologies with a factor four increase in the mass of the E-belt.